# Amplification of Interstellar Magnetic Fields and Turbulent Mixing by Supernova-Driven Turbulence Part II – The Role of Dynamical Chaos

By

## Dinshaw S. Balsara[1] and Jongsoo Kim[2]

(dbalsara@nd.edu, jskim@kasi.re.kr)

[1]Center for Astrophysics, Department of Physics, University of Notre Dame,

[2]Korea Space Science Institute

**Abstract**

In this paper we further advance the study of magnetic field amplification in the interstellar medium that was started in Balsara et al (2004, Paper I). We show that the flux growth rate is comparable to the rate of magnetic energy growth found in Paper I. We also demonstrate the role of intermittency in field amplification. The density shows a double-peaked PDF, consistent with the cooling curve that was used. The PDF of the magnetic field shows a high-end tail, providing a tell-tale signature of the operation of the small scale dynamo. The magnetic field strength correlates with the density as $|B| \sim \rho^{\alpha}$ with $\alpha = 0.386$. As a result, the field amplification takes place more vigorously in the lower temperature, denser gas. The Lagrangian chaos in the simulated turbulent flows is studied in substantial detail. It is shown that the stretching rate of material lines as well as the Lyapunov exponents can be used to gain important insights into the growth of magnetic field. The cancellation exponent for the small scale supernova-driven dynamo is derived and it is shown that constructive folding of field lines in the dynamo is very inefficient. We also show that our Lagrangian approach can yield actual measures of the turbulent diffusivity in the simulated ISM. The turbulent diffusivity provides insights into the mixing of elements from supernova ejecta on macroscopic scales. The high rates of line stretching in interstellar turbulence suggests that the eventual diffusion of elements at the molecular level is very efficient. Many of the diagnostics of turbulence that are presented here can be used to make direct connections between simulations and observations.

I) **Introduction**

The amplification of magnetic fields in proto-galactic environments, in the inter-cluster medium of galaxy clusters, in galaxies such as our own and in molecular clouds within our Galaxy is a topic of substantial interest in astrophysics. Examples of observations of magnetic activity in proto-galactic environments include observations of finite rotation measures in high-redshift quasars by Kronberg & Perry (1982), Perry, Watson & Kronberg (1993) and Kronberg (1995) and the observation of fields of a few µG in damped Ly-α systems at *z=2* by Wolfe, Lanzetta & Oren (1992). Kim, Kronberg and Tribble (1991), Vogt, Dolag and Enslin (2004) and Schuecker et al (2004) have made observations of magnetic fields in clusters. Molecular clouds within our own Galaxy probably contain supersonic, trans-Alfvenic turbulence, see Crutcher (1999), indicating that the magnetic fields are roughly close to their equipartition values. Observations of magnetic fields in our own Galaxy are being made with increasing frequency starting with the early work of Rand & Kulkarni (1989) and Beck et al (1996) and continuing with more detailed attempts to analyze the structure of the magnetic field in Minter & Spangler (1996), Beck (2001), Fosalba et al (2002) and Han, Ferriere & Manchester (2004). The results by Minter & Spangler (1996), Fosalba et al (2002) and Han, Ferriere & Manchester (2004) are most interesting because they show evidence for a break in the magnetic field spectrum. The spectral data shows evidence for a shallow spectrum for the large scale magnetic field and a steeper spectrum for small scale magnetic field. While the spectrum of the large scale magnetic field arises from the mean Galactic magnetic field, Minter & Spangler (1996) attribute the small scale magnetic field as arising from interstellar turbulence. Minter & Spangler (1996) find that the spectral index of the small scale magnetic field is consistent with Kolmogorov turbulence. The growth of large scale magnetic field probably takes place due to some form of mean field dynamo theory and is not the object of this study. The growth of small scale magnetic field may perhaps be attributed to the operation of a small scale dynamo and is being studied here.

In all of these objects, the large length scales and low viscosities of the system coupled with the ready availability of mechanisms for driving the turbulence make it very likely that turbulent flows establish themselves. Mechanisms that drive strong shocks into such systems are also easy to find and it is in fact quite likely that the strong shocks play a major role in sustaining the turbulence. For example, protogalactic environments and the environments of early quasars display a substantial amount of high mass star formation activity, quite like the starburst phenomenon observed at lower redshifts. It is, therefore, likely that supernovae from an early epoch of star formation might drive strong shocks through protogalactic environments. Churchill & Le Brun (1998) and Ledoux, Srianand & Petitjean (2002) find metallicities in damped Lyman $\alpha$ systems that are close to solar values, thus making the case for elevated rates of star formation in high redshift environments. In the case of clusters, the bow shocks and turbulent wakes of galaxies (Balsara, Livio & O'Dea 1994), the presence of jets ( Balsara & Norman 1992), galactic superwinds (Suchkov et al 1994) and cluster formation and merger (Miniati et al 2001) could drive turbulence in the inter-cluster gas and all these mechanisms are capable of driving strong, large-scale shocks in the gas. In the case of molecular clouds, jets and protostellar winds could be strong sources of shock-driven turbulence.

The best case for interstellar turbulence can be made by studying our own Galaxy. McKee & Ostriker (1977) drew on the work of Field, Goldsmith & Habig (1969) and Cox & Smith (1974) to propose a three-phase interstellar medium that was kept turbulent by supernova (SN) shocks and winds from OB associations. Scalo & Elmegreen (2004), Elmegreen & Scalo (2004) and Mac Low and Klessen (2004) provide recent reviews of the turbulent ISM. Balsara, Benjamin & Cox (2001; hereafter BBC01) have studied the interaction of three dimensional SN remnants with a turbulent, magnetized ISM. BBC01 found that the interaction of strong shocks with compressible turbulence and its resultant density fluctuations could be a strong source of helicity generation. Korpi et al. (1999) carried out low resolution simulations of SN-driven, magnetized interstellar turbulence. Kim, Balsara & Mac Low (2001) studied the energetics, structure and spectra of such a turbulent medium with substantially higher resolution simulations. Balsara & Kim (2004) drew on the RIEMANN framework for computational astrophysics described in Balsara (1998a,b, 2001, 2004) and Balsara & Spicer (1999a,b) to show the importance of good numerical magnetohydrodynamic (MHD) methods in enabling a faithful representation of magnetic field amplification in such SN-driven turbulence. Balsara et al (2004; hereafter Paper I) made a study of magnetic field amplification in such environments, finding that the helicity-generation mechanism from BBC01 was adequate to sustain field growth. Haugen, Brandenburg and Mee (2004) have simulated magnetic field amplification in an isothermal, compressible, shock-driven plasma. Their work complements our Paper I in demonstrating the importance of shocks in field amplification. Mac Low et al. (2005) have studied the thermodynamic nature of interstellar turbulence with SNe being the dominant energy input mechanism. Mac Low, Kim and Balsara (2005) have studied the role of supernova rates and magnetic fields in setting the hot gas filling fractions in SN-driven models of the ISM. Balsara and Kim (2005) have studied the turbulent diffusivity in SN-driven models of the ISM and the role of magnetic fields in quenching it. The simulation work cited above clearly indicates that the shock-driven interstellar turbulence is supersonic and that compressibility effects produce a range of phenomena in the ISM.

Mean field dynamo theory, Parker (1971), Steenbeck, Krause & Radler (1966) and Ruzmaikin, Shukurov & Sokoloff (1988) has been extensively applied to study the growth of large-scale magnetic fields in our turbulent Galactic ISM. Diamond, Hughes and Kim (2004) have presented a modern review of mean field dynamo theory. Ferrière (1993a, b, 1995, 1998) showed that SN explosions and superbubbles in the sheared and stratified Galactic ISM could provide one possible mechanism for helicity generation. Ferrière & Schmitt (2000) predict growth times for the mean Galactic field of about 1.8 Gyr for most of their models. Kulsrud & Anderson (1992) have questioned the applicability of mean field dynamo theory to high Reynolds number systems. While the growth of large scale magnetic fields probably lies within the purview of mean field dynamo theory, there are other ways of generating field more rapidly on smaller scales. We focus on the smaller scale growth of magnetic field in this paper. Kinematic methods for studying field growth go under the name of fast dynamos and have been catalogued in the text of Childress & Gilbert (1995, hereafter CG) and a more recent review by Galloway (2003). Several authors, Vainshtein & Zeldovich (1972), Otani (1993), Arnold & Korkina (1983), Galloway & Frisch (1986), Galanti, Pouquet & Sulem (1993), Lau & Finn (1993), Galloway & Proctor (1992), Galloway & O'Brian (1993) and Archontis (2000) have studied flows that lead to rapid amplification of magnetic field in the limit where the magnetic Reynolds number ($Re_m$) tends to infinity. The theory of fast dynamos focuses on magnetic field growth in this limit. While such a limit is unrealizable in astrophysical codes, the insights obtained from fast dynamo studies can be used to understand astrophysical simulations, as we illustrate in this paper. The demonstration that Lagrangian chaos in the underlying flow plays an important role in field amplification has also been made in Arnold & Korkina (1983), Galloway & Frisch (1986), Galanti, Pouquet & Sulem (1993), Klapper and Young (1995), Brandenburg, Klapper and Kurths (1995) and Galloway (2003). While Lagrangian chaos is not a necessary prerequisite for field growth, the previously mentioned papers indicate that the existence of such chaos usually leads to rapid growth of magnetic field. Kraichnan (1976a,b) had already studied dynamos in systems with strong kinetic helicity fluctuations of either sign where the mean kinetic helicity sums to zero. Galloway & Proctor (1992) were the first to show that Lagrangian chaos could lead to a fast dynamo even in flows without a net kinetic helicity. While the above demonstrations were kinematical, the fact that suitably chaotic driving can lead to field generation was shown via simulations by Zienecke, Politano & Pouquet (1999) for incompressible MHD and by Balsara (2000) for compressible MHD. Via the study of the turbulent ISM, Paper I provided an instantiation of a practical system, i.e. our Galaxy, where field was amplified rapidly, on order of the eddy turnover time. Paper I reconfirmed the scenario for kinetic helicity generation in BBC01, showing that the interaction of strong, supernova-driven shocks with a compressible, multiphase, turbulent ISM can generate large fluctuations in kinetic helicity of either sign. The forcing in Balsara et al (2004) was driven on a range of length scales by supernova shocks and the resulting magnetic field growth was also observed on all scales that were present in the computational domain. While such scales are small compared to the scale of the Galaxy, they are not at all small compared to say a globular cluster where a similar turbulent process may have operated at the time of galaxy-formation, see Saleh, Beers & Mathews (2004). Several analytic and incompressible numerical MHD studies have also appeared, Schekochihin et al (2004) and Lanotte et al (2004). The above authors have presented an interesting range of viewpoints on the growth of small scale magnetic fields, though they are all in agreement on the existence of a small scale dynamo.

The simulations in Paper I can be viewed as an instantiation of a small scale dynamo in the Galaxy with an astrophysically motivated driving, using shock-capturing methods. The approaches of Schekochihin et al (2004) and Lanotte et al (2004) can be viewed as idealized simulations of small scale dynamos. Despite the parallel goals, there are several interesting points of difference between the two approaches that are useful to catalogue:

a) The "white in time" forcing that has been used in many of the studies of small scale dynamos imply that the turbulent forcing has a zero coherency time. Real Galactic flows are driven by the expansion of SNRs which are observable as coherent structures for about 80 thousand years which is a good fraction of the 0.6 Myr turbulent coherence time that we deduce from our simulations in this paper. Thus the utility of the "white in time" forcing might be worth a more detailed examination.

b) The SNe drive the turbulence on a wide range of length scales. Some of the driven turbulence studies of small scale dynamos only drive the turbulence over a narrow range of length scales.

c) The effect of compressibility has been ignored in several of the studies of small scale dynamos. However, density and magnetic field strength are found to correlate in MHD turbulence. This is a correlation that we showed visually in Fig. 9 of Paper I and we will demonstrate it again more formally in this paper.

d) The role of strong shocks in producing vorticity and helicity when they interact with density discontinuities is ignored. Samtaney and Zabusky (1994) have shown that the interaction of shocks with density discontinuities can be an important source of vorticity and BBC01 have shown that the inclusion of turbulence leads to helicity-generation.

e) Turbulent forcing via SNe shocks can be strong enough to overwhelm the magnetic field, even at saturation.

To their advantage, the incompressible approaches to studying the small scale dynamo can probe a range of magnetic Prandtl numbers and points a) and b) above are indeed easy to overcome. The compressible, shock-capturing approaches for studying the small scale dynamo operate with effective Prandtl numbers that are close to unity but they do address points c), d) and e) above. Thus there are benefits to either approach and both yield valuable perspectives.

A common theme that runs through fast dynamos as well as small scale dynamos is the importance of studying Lagrangian chaos in order to understand field growth. CG alternatively refer to their fast dynamos as stretch, twist and fold (STF) dynamos and go on to emphasize the importance of stretching and constructive folding in field growth. Measures of line stretching, Lyapunov exponents and constructive folding are shown to yield important perspective on field growth. An examination of those measures in the context of the simulations in Paper I yields interesting perspectives and we report on such a study in this paper. Because these concepts may be new to the astronomy community, we also explain these concepts in detail and provide pointers to the other literature.

Implicit in any study of the mean field dynamo is the issue of turbulent diffusivity of the ISM. The turbulent diffusivity of vector fields is a difficult topic that has not yet been well-resolved in the literature. Impressive progress has, however, been made by several authors, see for example Gruzinov & Diamond (1994, 1995) or Kim & Diamond (2001, 2003). The turbulent diffusivity of a passive scalar is, on the other hand, a problem where some resolution has been obtained and recent reviews have been provided by Shraiman & Siggia (2000), Warhaft (2000) and Falkovich, Gawedzki & Vergassola (2001). As a result, researchers often study the turbulent diffusivity of a passive scalar as a substitute for the study of turbulent diffusivity of vector fields, see Childress & Soward (1989) for example. Under an assumption of scale separation in the ISM, Steenbeck, Krause and Radler (1966) obtain tensor expressions for the turbulent diffusivity. The problem, however, is that scale separation may not always be a good approximation in the Galactic ISM, especially in light of the prior observation that SNe and superbubbles can drive the ISM over a broad range of length scales. The issue is brought home by the realization that the supernovae and OB associations drive the turbulence that is needed for the small scale dynamo and are also invoked to produce the mean helicity that is needed for the mean field dynamo on larger scales, see Ferriere (1993a, b). The simulations presented here can be used to provide measures of the turbulent diffusivity of the ISM, which can be used to obtain quantitative measures for the diffusion of scalar fields in the ISM and also perhaps be used to make a more quantitative analysis of the diffusivity of vector fields in mean field dynamos.

Measuring the turbulent diffusivity of the ISM is also important because it plays an important role in the study of Galactic chemical evolution and mixing. In our present ISM the O/H ratio is almost homogeneous while the D/H ratio is not, see Jenkins et al (1999), Sonneborn et al (2000) and York (2002) for data relating to the Interstellar Medium Absorption Profile Spectrograph (IMAPS). The Pleiades cluster was presumably born in a turbulent giant molecular cloud (GMC) and the stars in that cluster show nearly identical metal abundance, see Wilden et al (2002). This again suggests that turbulent diffusivity in turbulent GMCs is very effective at producing chemical homogeneity. In a study of low metallicity halo stars with -3.4 < [Fe/H] < -2.2, Arnone et al (2005) find that the [Mg/Fe] ratio is very homogeneous, with fluctuations that are less than 0.2 dex. The result is surprising because it stands in contradistinction to the one-zone chemical evolution models which predict a scatter of 1 dex in the [Mg/Fe] ratio at those metallicities. Arnone et al (2005) argue that the proto-galactic environment is well-mixed. Mathews & Cowan (1990) point out the importance of r-product elements such as Eu in tracing SN progenitor stars in specific mass ranges. Such elements might, therefore, give us an even better handle on turbulent mixing. A clear, computationally driven understanding of turbulent diffusivities in such environments would provide a better perspective on the problem of mixing. The problem can be split into two parts. First, we are interested in the rate at which interstellar turbulence diffuses metal-rich SN-ejecta on macroscopic scales. Second, as pointed out by Oey (2003), it is also important to pay attention to processes by which the metal-rich ejecta get stretched into long, thin structures. Molecular diffusivity can operate on the short transverse length scales of such narrow structures to produce true mixing at the molecular level. In this paper we build diagnostics that allow us to quantify each of those two parts. We also point out that measures of the turbulent diffusivity can be used in conjunction with the high rates of mixing predicted in Arnone et al (2005) to set a timescale for the rapidity of star formation in the Galactic halo.

In Paper I we studied some aspects of the compressible small scale dynamo. However, the issues of flux growth and the importance of Lagrangian chaos in the growth of magnetic fields were deferred to another paper. We take up these issues in the present paper. Since the numerical models were already described in Paper I we will not repeat the description here. We mostly focus on the $256^3$ zone run reported in Paper I but the other runs from Paper I have also been used for confirmation. In Section II we study the growth of flux. In Section III we study the effective turbulent diffusivity of the ISM. In Section IV we study the role of dynamical chaos in amplifying magnetic fields. In Section V we offer some discussions and conclusions.

**II) Flux Growth and Magnetic Field Structure**

In the first part of this section we first study the growth of magnetic flux. In Paper I we did not correlate the dynamo-generated magnetic field strength and structure with various aspects of the flow. We do that in the second part of this section.

**II.a) The Growth of Magnetic Flux**

Early practitioners of fast dynamo theory were concerned about their field amplification indeed being a dynamo, see Finn and Ott (1988a, b). It was theoretically possible that the flows being considered could shred the magnetic field into successively thinner sheets or filaments without producing a net magnetic flux. Such a field would have shown an increasing magnetic energy without showing an increasing net flux. Finn & Ott (1988a, b) showed that such was not the case for fast dynamos. It was realized that the concept of constructive folding of an STF dynamo in three dimensions plays an important role in flux growth. Realize, therefore, that a magnetic field that is folded in two dimensions cannot but achieve a structure that is folded into successively narrower, oppositely-oriented bands. Three dimensional effects offer a way out, see CG. Simple stretching, twisting and folding in three dimensions does not always guarantee flux growth. However, if the twisting and folding work constructively with each other in three dimensions then they will produce a net growth of magnetic flux, see Fig. 1.3 in CG for a visual example of constructive folding. Not every kind of three dimensional turbulent flow will demonstrate constructive folding, and it is very likely that many of the motions in a turbulent flow cause destructive folding of the magnetic field. However, three dimensional flows with large excursions in local kinetic helicity are likely to show some net constructive folding. See BBC01 or Paper I for the helicity-generation in our simulations. Since the helicity in our simulations is mirror symmetric, we do not expect a global magnetic flux to build up. Also, since it is time-dependent and random in all three dimensions we do not expect it to persistently build up a net flux in a specified direction in a given plane. However, on average, we do expect to see growth in the magnitude of the flux crossing a given plane in either direction.

To measure the magnetic flux, we set up three $64^2$ zone planes in the center of our computational domain with normal vectors pointing in the x, y and z-directions and measured flux through them as a function of time. Fig. 1 shows the fluxes $\Phi_x$, $\Phi_y$ and $\Phi_z$ through those planes as a function of time. It is easy to see that there are several episodes of flux reversal. However, on average, we do see a net growth of magnetic flux. Using a 5Myr sliding window, we averaged the growth rate of the maximum of the three fluxes and plotted the growth rate as a function of time in Fig. 2. The episodes of rapid flux destruction and regeneration in Fig. 1 correspond to large negative and positive growth rates respectively in Fig. 2. The time-averaged growth rate is shown by the solid line in Fig. 2. The dotted and dashed lines show the time-averaged growth rate in the positive and negative parts of the curve in Fig. 2 respectively. We see that the mean growth rate for magnetic flux, $\gamma_0$, for most of the time in the simulation is given by $\gamma_0 = (20.3\text{Myr})^{-1}$. This is roughly comparable to the growth rate of $(10\text{Myr})^{-1}$ for the net magnetic energy that we found in Paper I. We also see that the time-averaged growth rate of just the positive parts of the curve in Fig. 2 is considerably larger than $\gamma_0$ showing that there are rapid, prolonged episodes of flux growth. The absolute value of the time-averaged growth rate of just the negative parts of the curve in Fig. 2 behaves similarly, showing that intermittent, localized flux destruction plays an important part in dynamo growth. However, we see that from Figs. 1 and 2 that the episodes of rapid destruction of magnetic flux are always followed by even more vigorous and prolonged episodes of flux regeneration.

In a time-dependent flow we do not consider the same patch of fluid because the fluid is not tracked in a Lagrangian fashion. It is, nevertheless, useful to ask whether we can find some evidence of local intermittency in the flow and the magnetic field as a way of explaining the rapid flux destruction and regeneration? To that end, we considered the $64^3$ zone region that contains the above-mentioned three $64^2$ zone planes in the center of our computational domain. Within that region we measured the kurtosis in the magnetic field and flow variables as given by the ratios $\frac{\langle B^4 \rangle^{\frac{1}{2}}}{\langle B^2 \rangle}$, $\frac{\langle B^6 \rangle^{\frac{1}{3}}}{\langle B^2 \rangle}$ and $\frac{\langle v^4 \rangle^{\frac{1}{2}}}{\langle v^2 \rangle}$ and plotted them as a function of time in Fig. 3a. Fig. 3b shows $\frac{\langle B^2 \rangle}{\langle \mathbf{B} \rangle^2}$ as a function of time. The kurtosis of a flow variable measures the inverse filling fraction of that flow variable and is, therefore, a good indicator of the local level of intermittency in the flow. To take the ratio $\frac{\langle B^4 \rangle^{\frac{1}{2}}}{\langle B^2 \rangle}$ as an example, a mean large-scale field would contribute significantly to the denominator while a local sub-region that represents a substantial excess over the mean field would contribute strongly to the numerator. Thus $\frac{\langle B^2 \rangle}{\langle B^4 \rangle^{\frac{1}{2}}}$ measures the fractional volume that has magnetic field values that are in substantial excess of the mean field. The variable $\frac{\langle B^4 \rangle^{\frac{1}{2}}}{\langle B^2 \rangle}$ is, therefore, a good indicator of the intermittency in the magnetic field. We see from Fig. 3a that the kurtosis of the magnetic field is not constant. This indicates that intense smaller scale magnetic structures keep forming and dissipating in the magnetic field. This result is consistent with our expectation for a dynamo in a high $Re_m$ flow. The kurtosis of the velocity field also fluctuates, however, those fluctuations are not as pronounced as the fluctuations in the magnetic field because the velocity field is not going through a process of dynamo growth. The magnetic and velocity fields in our simulation are thus found to be intermittent, with the magnetic field being very strongly intermittent. As time evolves, the intermittency in the magnetic field decreases because the field is growing and some portions of the flow have field strengths that are close to equipartition, as will be seen in Fig. 4. Because we have not used Lagrangian coordinates and also because we have used a rather large volume of space, we see from Fig. 1 and Fig. 3a that the concordance between epochs of high intermittency and epochs of rapid flux destruction is not exact. We do, however, see that episodes of high intermittency show a rough correlation with episodes of rapid flux destruction or regeneration. The variable $\frac{\langle B^2 \rangle}{\langle \mathbf{B} \rangle^2}$ presented in Fig. 3b is also astrophysically interesting because synchrotron emission measure tracks $\langle B^2 \rangle$ while the pulsar rotation measure tracks $\langle \mathbf{B} \rangle$. As time increases, we see an overall increasing trend in the value of this ratio. Because we do not follow the dynamo to saturation, we cannot predict the saturation value. Our simulations do not include the processes that lead to the formation of mean magnetic field so the diagnostic we produce here is not directly comparable to Galactic values, except perhaps on the smallest scales. However, Fig. 3b shows the utility of simulated variables such as $\frac{\langle B^2 \rangle}{\langle \mathbf{B} \rangle^2}$ in making connection with observations.

**II.b) The Structure of the Dynamo-Generated Magnetic Field**

The structure of the resulting dynamo-generated magnetic field is also a topic of great interest. It is very interesting to relate that to the structure of the density variable, since the density is a direct tracer of compressibility effects in the flow. It is also interesting to try and correlate the field strengths to the densities in the flow. We do that in this sub-section with one caveat. Our simulations have been carried to an epoch where quasi-linear saturation of the dynamo has just set in with the ratio of magnetic pressure to the gas pressure being about 0.01 at the end of the simulation. It is possible that the small scale dynamo field in the kinematic and quasi-linear regimes might look very much like the small scale dynamo field in the saturated regime. However, the saturated small scale dynamo field could also produce PDFs that are slightly different from PDFs of the magnetic field in the kinematic regime, see Schekochihin et al (2004) for an example of that. Log-normal PDFs for the magnetic field strength in fast dynamos were first predicted by Finn & Ott (1988a,b) using maps and later on by Lau and Finn (1993) using incompressible ABC flows.

Fig. 4 shows the PDFs of the absolute magnitude of the magnetic field, $|\mathbf{B}|$, at different epochs in the dynamo simulation. Fig. 5 shows the corresponding PDFs of the density at the same epochs in the dynamo simulation. We see that the magnetic field strengths in Fig. 4 cannot be fit by a log-normal distribution. The density in Figs. 5 shows a double-peaked distribution. Such a double-peaked distribution arises because our cooling curve is obtained from the work of MacDonald & Bailey (1981) and does not incorporate some of the molecular cooling terms that were shown to be important in Wolfire et al (1995, 2003). As a result of our choice of cooling, the simulations produce the hot and warm phases of the ISM but do not have the requisite physics for producing the third, molecular phase. Incorporation of that further bit of physics will be done in future work. Such molecular cooling terms have been included in Gazol et al (2001) who, however, exclude the SN and, therefore, produce the molecular and warm phases but do not produce the hot phase. The PDFs for the density saturate within the first 5 Myr of the simulation, showing that the density does not evolve after a while. Fig. 4 shows that the PDFs for the magnetic field strength keep shifting to the right as a function of time, showing the growth of magnetic energy in the small scale compressible dynamo. The presence of a high-end tail in the PDFs of the magnetic field can be seen at all epochs in Fig. 4. The high-end tail in Fig. 4 arises from compressibility effects which can produce locally enhanced densities and, therefore, locally enhanced field strengths. The high-end tail in the magnetic field in Fig 4 also precludes a log-normal fit to the PDFs of the magnetic field. The long tails at the high end in the PDFs in Fig. 4 for the magnetic field strength show us that local regions with high values of the magnetic field strength can be produced by the small scale dynamo. These local regions would have rather small filling factors, thus explaining the high intermittency in the magnetic field that was observed in Fig. 3a. At a time of 20Myr we see from Fig. 4a that very small portions of the computational domain have magnetic fields that are in excess of 6.5 μG, which is the equipartition value of the magnetic field in our simulated ISM. Note too that our SNe rate is eight times larger than the Galactic rate, accounting for the larger equipartition value. We, therefore, realize that small portions of our computational domain may already have reached equipartition, though that fraction is very small even towards the end of the simulation at 40 Myr. This approach to equipartition also explains why the kurtosis in Fig. 3a showed decreasing fluctuations towards the end of the simulation.

Fig. 6 shows the scatter plot of the logarithm of the magnitude of the magnetic field v/s the logarithm of the density at a time that corresponds to the end of the dynamo simulation. The hot gas is shown in red and has $\log_{10} T > 5.5$ where "T" is the temperature. The transitional gas is shown in green and has $3.9 \leq \log_{10} T \leq 5.5$ and is the gas that is not likely to be in thermal equilibrium equilibrium. The warm gas is shown in blue and has $\log_{10} T < 3.9$. We find a very strong positive correlation between the magnetic field and the density in Fig 6, showing that compressibility effects do play a role in the generation of magnetic fields in the compressible dynamo. Fig. 6 shows us that a majority of the field amplification takes place in the denser gas. Since the denser gas in Fig. 6 is likely to come from the warm and transitional phases while the under-dense gas is likely to correspond to the hot phase, we do not consider it optimal to try and fit a single power law through the entire data set in Fig. 6. However, we realize that such information may be desirable to astronomers so we, nevertheless, provide such information. Our best fit to the relation between the magnetic field strength and the density goes as $|B| \sim \rho^\alpha$ with $\alpha = 0.386$ over the entire data set in Fig. 6. The variation of magnetic field with density is more likely to be measured only in the warm and transitional gas phases. Thus we make another fit to just the warm and transitional phase data in Fig. 6 and find $|B| \sim \rho^\alpha$ with $\alpha = 0.369$. It is also worthwhile pointing out that Fig. 6 shows a trend with a very large scatter. As a result, observations that are similar to those presented by Heiles & Troland (2003) would have to build up substantial amounts of statistics to produce reliable B-n relations. We have also used the other simulations from Paper I to provide added confirmation of the positive correlation between the magnetic field strength and the density. We point out that the magnetic field is still unsaturated by the end of the simulation reported here. However, similar positive correlations between magnetic field strength and the density have also been found in the saturated simulations that were reported in Mac Low et al (2005). Passot et al (2003) have also tried to correlate the magnetic field strength to the density but because of their use of an isothermal equation of state a direct comparison is not possible.

**III) The Effective Turbulent Diffusivity of the ISM**

The early literature by Batchellor (1949) has provided a mixing length theory for turbulent diffusion. de Avillez & Mac Low (2002) tried to analyze the diffusion of scalar fields in the ISM but their ideas were based on coloring squares and, therefore, may not easily translate into quantitative measures of turbulent diffusivity. Drummond, Duane & Horgan (1984) have developed Lagrangian strategies for measuring the diffusivity of a scalar field in a turbulent fluid. Such a Lagrangian approach is also underscored in Frisch (1995). Klessen & Lin (2003) pointed out the importance of removing the mean drift in measures of turbulent diffusion. The method of Drummond, Duane & Horgan (1984) and Klessen & Lin (2003) consists of tracking a set of Lagrangian particles in the flow. The numerically measured specific scalar diffusivity is then given by:

$$\eta_{\text{turb}} = \frac{1}{3} \frac{d}{dt} \left\langle \left[ \{\mathbf{r}(t) - \langle \mathbf{r}(t) \rangle\} - \{\mathbf{r}(t_0) - \langle \mathbf{r}(t_0) \rangle\} \right]^2 \right\rangle$$

$$\approx \frac{1}{3} \frac{1}{(t - t_0)} \left\langle \left[ \{\mathbf{r}(t) - \langle \mathbf{r}(t) \rangle\} - \{\mathbf{r}(t_0) - \langle \mathbf{r}(t_0) \rangle\} \right]^2 \right\rangle$$

Notice that the above equations remove the mean drift of the particles. While the above expression is to be evaluated in the limit $t - t_0 \to \infty$, in practice it can be evaluated for a time that is larger than the coherence time in the simulation. To obtain the turbulent diffusivity, we set up a $64^3$ mesh of particles with an initial distance of one zone size between them in all three directions. The diffusivity is measured by allowing the particles to evolve by using the local velocity of the flow. An integration scheme that was fourth order accurate in its spatial interpolation and second order accurate in time was used to evolve the particles. We display the variation of $\eta_{turb}$ as a function of time in Fig. 7. From Fig. 7 we find that $\eta_{turb}$ initially increases with time but asymptotically tends to a value of $5.7 \times 10^{26}$ cm$^2$ / sec. Fig. 7 shows that it takes a time of 0.6 Myr before $\eta_{turb}$ approaches its asymptotic value. Since the swarm of $64^3$ particles was initialized on a cube with sides that were 50 pc long, we, therefore, deduce that the coherence time of the turbulence on scales of 50 pc is given by $\tau_{coherence} = 0.6$ Myr. Before $\tau_{coherence}$, the particles move ballistically on the eddies that carry them. Thus over a short duration of time that is less that $\tau_{coherence}$, for any particle "i" we have $\mathbf{r}_i(t) \approx \mathbf{r}_i(t_0) + \mathbf{v}_i (t - t_0)$. Here $\mathbf{r}_i(t)$ is the position of the i$^{th}$ particle at a time "t" and $\mathbf{v}_i$ is the local velocity of the eddy that carries the particle. Past $\tau_{coherence}$, the particles begin to carry out their random walk on an ensemble of turbulent eddies that have become uncorrelated, yielding a constant value for $\eta_{turb}$. The density-weighted r.m.s. velocity in the simulation was measured to be $v_{rms} = 12.75$ Km/sec. This value is higher than the r.m.s. velocity measured in our current ISM but is consistent with the fact that the SN-driving is substantially more vigorous in the simulations reported here. We observe that $\eta_{turb}$ agrees with $v_{rms}^2 \tau_{coherence}$ to within an order of magnitude, showing that it is possible to interpret our detailed measurement of $\eta_{turb}$ by using a mixing length theory. However, to get precise, quantitative measurements of $\eta_{turb}$ it is important to use the Lagrangian approach presented here.

The technique for measuring turbulent diffusivities in the above paragraph is interesting because it can now be applied to entire classes of simulations of the ISM, making it possible to get numerical measures of the turbulent diffusivity and its dependence on supernova rate and ISM parameters. Such an assessment will be reported in Balsara & Kim (2005).

**IV) The Role of Lagrangian Chaos in Magnetic Field Growth**

The evolution of a fast dynamo requires a STF or stretch, fold and shear mechanism to operate. By showing the flux growth in Section II.a we showed that the flows in our problem do possibly have constructive folding. In Paper I we showed that helicity is naturally generated in these flows, providing a mechanism that twists the field lines. Neither the stretching of field lines nor the constructive folding of field lines has yet been illustrated for the flows in our problem. Field line stretching is very important in producing a volume-filling, amplified field. To see the importance of stretching, imagine a loop of magnetic field that is twisted and folded as shown in Fig. 1.3 of CG but with the stretching step being omitted. Successive applications of twisting and folding (without stretching) would result in a strong field, but one which is concentrated in a very small volume, resulting in no net increase of magnetic energy. To see the importance of constructive folding, realize that a twist of $180^o$ in the loop of magnetic field is needed before the folding step to achieve perfect amplification of the field. Any twist greater than $90^o$ before the folding step would have resulted in some, less than perfect, amplification of the field. Twists of less than $90^o$ before the folding step would have resulted in destruction of magnetic field. Constructive folding measures the extent to which the twisting and folding are conducive to field amplification. In the following three sub-sections we examine three increasingly sophisticated measures of Lagrangian chaos and their relation to field growth via field line stretching and constructive folding.

**IV.a) Line Stretching**

In the limit of ideal MHD we know that the magnetic field lines are materially advected by the flow velocity. As a turbulent flow evolves, it stretches the field lines embedded within it. The measure of field growth that we explore in this sub-section is related to such a stretching process, which in turn is related to the STF mechanism in a fast dynamo. To measure field line stretching, we initialized several points along three straight line segments that were oriented along the x,y and z-directions in the flow. The segments had an initial length of 50pc and were placed in the center of our computational domain. Such line segments can be thought of as representing a section of a magnetic field line. The points along those line segments were integrated in time with the local Lagrangian velocities of the flow. The spatial interpolation of the velocities to the points that make up each segment was done using a four-point Lagrange polynomial fit. The temporal integration of points was done using a second order accurate predictor-corrector method. If the distance between any two points along any of the segments became larger than two zone widths, new points were adaptively inserted between the original points using an arithmetic interpolation algorithm. We kept track of the periodic wrap-around of points within the computational domain and the appropriate periods were put back in when making measures of line stretching. While each of the line segments initially started off with 64 points, in about one Myr the above-mentioned adaptive stretching algorithm caused the lines to have as many as $10^6$ points. When measuring line stretching we had to keep track of the ordered numbering of the points that make up the line segments. This forced a serialization of the line stretching algorithm described above. As a result, it was not possible to follow line stretching for very long durations of time. However, the results that will be presented show that line stretching produces strongly folded structures on smaller scales and we expect that our time of one Myr exceeds the coherence time for eddies on those scales. In Section III we have shown that the coherence time for eddies is 0.6 Myr.

Fig. 8a shows the evolution of the lengths of those line segments in the flow as a function of time, starting at a time of 20Myr in the simulation. The two thick dashed lines in Fig. 8a mark out line stretching rates of $(0.10\text{Myr})^{-1}$ and $(0.11\text{Myr})^{-1}$. The line stretching shows an initial growth rate of $(0.11\text{Myr})^{-1}$ but the asymptotic growth rate can be seen to be $(0.10\text{Myr})^{-1}$. Fig. 8b shows the rms value of the radii of the points that make up each segment (with the mean subtracted off) as a function of time. Fig. 8c shows the segment that was originally oriented along the y-axis at a time of 20.08Myr, i.e. 0.08Myr after the lines were evolved. Fig. 8d and Fig. 8e show the same segment at times of 20.4Myr and 20.8Myr. Fig. 8a shows us that the line segments grow exponentially, indicating that the magnetic field lines are also stretched exponentially by the flows that develop in our simulations. Exponential stretching of magnetic field lines is an essential prerequisite for the existence of a fast dynamo, as was shown by Vishik (1989). Vishik (1989) also finds that the rate of line stretching sets an upper bound to the rate of growth in a fast dynamo, a finding that is corroborated by our own measurement of the growth rate for line stretching in our simulations. Note though that Vishik's theorem is strictly applicable only to flows without discontinuities, which excludes all shock-dominated astrophysical flows. Since strong shocks only fill a modest fraction of the volume in our simulations, Vishik's theorem can still give us some insight into magnetic field evolution in the parts of the flow that do not have shocks. Fig. 8b shows us that the rms radius associated with each of the line segments is substantially smaller than the length of the line segments, indicating that the line segments are strongly folded. Fig. 8c shows the line segment a very short time after its Lagrangian motion was started. We see from Fig. 8c that the line has already formed a loop. The loop-like structure illustrates the effect of the helical motions that naturally arise when SNe shocks interact with turbulence, as first shown in BBC01. In Paper I we also analyzed the compressive and solenoidal parts of the velocity spectrum and found that the solenoidal component dominates, providing a dynamical motivation for the loop-formation that we see in Fig. 8c. As time progresses, the line gets twisted into many more loops, as shown in Fig. 8d. The loop-like structures that we see in Figs. 8c and 8d are very indicative of Lagrangian motion in a velocity field with a dominant solenoidal component. Fig. 8e shows the original line segment at a time of 20.8Myr. The progression from Figs. 8c to 8d to 8e makes the exponential line stretching from Fig. 8a graphically evident. We also see from Figs. 8c, 8d and 8e that the line is strongly folded even as it is stretched exponentially. It is because of this high rate of folding that the line length can be as large as $10^3$ Kpc within 1Myr, while the r.m.s. distribution of the points that make up the line segments can be smaller than 0.1 Kpc, see Figs. 8a and 8b. An animation of field line stretching shows that the rate of line stretching is fairly uniform in time. However, there are epochs when the line length can increase in rapid bursts. Such epochs can also be discerned from the two slight kinks that can be seen in Fig. 8a. Examination of the placement of successive generations of SNe explosions revealed that such bursty evolution takes place when a supernova explodes in the shell left behind by a prior supernova explosion.

It has been suggested, see Parker (1979) and Kulsrud & Anderson (1992), that the amplification of the Galactic magnetic field to equipartition in a Hubble time requires an effective reconnection rate that is much higher than the reconnection rate suggested by atomic values. Yokoyama & Shibata (1994) have suggested that anomalous resistivity might be one possible way of increasing the reconnection rate. The Magnetic Reconnection Experiment (MRX) has presented direct evidence for turbulent plasma processes that might lead to an anomalous resistivity, see Terry et al (2003). Figs. 8c to 8e show that material lines are strongly folded by the turbulent flows. This further suggests that magnetic field lines do the same. Such folding is very likely to bring oppositely oriented magnetic field lines into close contact, thus providing an alternative way of increasing the reconnection rate. Since much of the turbulence is mediated by strong shocks, it is quite possible that some of the reconnection is of the faster Petschek (1964) variety rather than the slower Sweet-Parker variety, see Sweet (1958) and Parker (1957). As argued in BBC01 and Lazarian & Vishniac (2000), an elevated reconnection rate would raise the possibility that reconnection may be an alternative source of heat for the ISM.

It is also interesting to want to know the conditions in a typical ISM that might produce flows with exponential line stretching. In Paper I we showed that a SN rate that is too high might result in thermal runaway thus bringing the small scale dynamo growth to an end. It is also possible that too low a SN rate would result in too little line stretching, bringing the small scale dynamo growth to an end. A very strong mean magnetic field or a very dense ISM can also impede the expansion of SNR shells resulting in a diminished rate of line stretching. This effect can be taken to be an analogue of $\alpha$-quenching, as it applies to the ISM. The exact conditions in the ISM for which line stretching is either diminished or enhanced will be studied thoroughly in a later paper, see Balsara & Kim (2005). It is, however, clearly demonstrated in this paper that line stretching in rather realistic interstellar flows plays an important role in the small scale dynamo and should be studied further.

In an idealized, maximally helical, chaotic flow, CG showed that the timescale for line stretching is comparable to the timescale for the growth of magnetic field. We see here that our growth rate for line stretching is much faster than our growth rate for the magnetic field. This indicates that while interstellar flows are capable of fast stretching, the field growth is not as fast as the line stretching because interstellar flows do not have good constructive folding. This would seem only natural in a randomly driven flow where no special effort has been made to fold the magnetic field lines at just the right rate to produce rapid field growth. Indeed, when simulating the Galactic ISM we cannot set up the problem to have just the right rates of constructive folding as was done in the idealized flows studied in CG and references therein. Even in the case of the carefully constructed, idealized, fast dynamo flows discussed in CG it was found that the folding was not always perfectly constructive.

The data presented in Figs. 8a to 8e can also be interpreted in an alternative way that yields important insights into turbulent mixing. It has been realized that turbulent mixing of metals in the ejecta of SNe down to the molecular level requires that the ejecta be stretched out into thin, narrow structures, at which point molecular diffusivity can produce a homogeneous mixture of elements, see Oey (2003). However, in the past, we did not have quantitative tools with which we could analyze the rate at which thin structures are formed in interstellar turbulence. Bulk turbulent diffusivity, based perhaps on a mixing length approximation from Batchelor (1949), does not furnish a complete understanding of this process. This is because the energy-bearing length scales of the turbulence are several orders of magnitude larger than the scales on which molecular diffusion operates. The line stretching can, on the other hand, give us a measure of the rapidity with which thin structures are formed in the flow. Consider a line segment with initial length $l_0$ that is stretched exponentially in time so that its length $l(t)$ at a later time "t" is given by:

$l(t) = l_0 \ e^{t/\tau_l}$

From Fig. 8a we see that $\tau_l = 0.10 \text{Myr}$. Imagine a cylinder with diameter $d_0$ around this line segment. Such a cylinder can, for example, be thought of as representing a volume that has a higher than average concentration of metallicity. Under an incompressible assumption, which is certainly not fully justified here, we can say that the diameter $d(t)$ at a later time "t" satisfies the relationship $l(t)\, d^2(t) = l_0\, d_0^2$ so that we have

$$d(t) = d_0\, e^{-t/(2\tau_l)}$$

The above formula gives us the rate at which thin, narrow structures are formed in a turbulent medium that has exponential line stretching. Thus for our reported simulation, narrow structures can grow narrower at an exponential rate and the characteristic time for that process is given by 0.20Myr. This characteristic time is much smaller than the estimated 100 Myr in Oey (2003) showing that a detailed examination of line stretching reveals that the mixing of elements down to the molecular level can proceed much faster than the estimates used in one-zone models for Galactic chemical evolution.

**IV.b) Lyapunov Exponents**

Numerous authors have found that the existence of Lagrangian chaos in a flow is a strong indicator that the flow might drive a dynamo. In the Introduction we have catalogued several of these studies. Zeinecke, Politano & Pouquet (1999) have shown that the Lyapunov exponents that arise naturally in a study of Lagrangian chaos can be used to good effect in understanding dynamical simulations of field amplification. The ensuing description of Lagrangian chaos has been detailed in Ott (1993) but we provide it below in a way that makes better contact with the MHD problem. To obtain the Lyapunov exponents, we set up a $64^3$ mesh of particles with an initial distance of one zone size between them in all three directions. Let the particles be labeled by indices (i,j,k) and let their three dimensional coordinates at a time "t" can be given by $x_1$ (i,j,k,t), $x_2$ (i,j,k,t) and $x_3$ (i,j,k,t). The particles are allowed to evolve in time using the local

Lagrangian velocity of the flow. Then for any later time we can construct the displacement gradient matrix:-

$$M(t) = \begin{bmatrix} \dfrac{x_1(i+1,j,k,t) - x_1(i,j,k,t)}{x_1(i+1,j,k,0) - x_1(i,j,k,0)} & \dfrac{x_1(i,j+1,k,t) - x_1(i,j,k,t)}{x_2(i,j+1,k,0) - x_2(i,j,k,0)} & \dfrac{x_1(i,j,k+1,t) - x_1(i,j,k,t)}{x_3(i,j,k+1,0) - x_3(i,j,k,0)} \\ \dfrac{x_2(i+1,j,k,t) - x_2(i,j,k,t)}{x_1(i+1,j,k,0) - x_1(i,j,k,0)} & \dfrac{x_2(i,j+1,k,t) - x_2(i,j,k,t)}{x_2(i,j+1,k,0) - x_2(i,j,k,0)} & \dfrac{x_2(i,j,k+1,t) - x_2(i,j,k,t)}{x_3(i,j,k+1,0) - x_3(i,j,k,0)} \\ \dfrac{x_3(i+1,j,k,t) - x_3(i,j,k,t)}{x_1(i+1,j,k,0) - x_1(i,j,k,0)} & \dfrac{x_3(i,j+1,k,t) - x_3(i,j,k,t)}{x_2(i,j+1,k,0) - x_2(i,j,k,0)} & \dfrac{x_3(i,j,k+1,t) - x_3(i,j,k,t)}{x_3(i,j,k+1,0) - x_3(i,j,k,0)} \end{bmatrix}$$

The displacement gradient matrix enables us to relate the components of an infinitesimal, material line segment ($\delta x^1(t)$, $\delta x^2(t)$, $\delta x^3(t)$) at a time=t to its components at a slightly later time=t+$\Delta$t given by ($\delta x^1(t+\Delta t)$, $\delta x^2(t+\Delta t)$, $\delta x^3(t+\Delta t)$) using the following matrix equation:

$$\delta x^i(t+\Delta t) = \sum_{j=1}^{3} M_{ij} \, \delta x^j(t).$$

This is useful because, in the limit of ideal MHD, the magnetic field components ($B^1(t)$, $B^2(t)$, $B^3(t)$) that are advected by the flow also evolve according to the Cauchy equation as follows:

$$B^i(t+\Delta t) = \sum_{j=1}^{3} M_{ij} \, B^j(t).$$

The close similarity between the above two equations shows us that studying the properties of the displacement gradient matrix can give us important insights into the evolution of the magnetic field. We construct the positive symmetric matrix $\dfrac{1}{2t} \ln [M^T(t) M(t)]$ and find its ordered eigenvalues $\lambda_1 > \lambda_2 > \lambda_3$ and its ordered, orthogonal eigenvectors $\mathbf{e}_1$, $\mathbf{e}_2$, $\mathbf{e}_3$. The eigenvalues are called the Lyapunov exponents and the eigenvectors are referred to as the Lyapunov eigenvectors. We can evaluate the displacement gradient matrix over short intervals of time to get instantaneous Lyapunov eigenvalues and eigenvectors or over longer intervals of time that are comparable to a coherence time in order to get cumulative Lyapunov eigenvalues and eigenvectors. In the case of incompressible flows, the Lyapunov exponents sum to zero. For compressible flows it is possible to have a SN shell that is expanding in all three directions and, therefore, has all positive Lyapunov exponents. The largest positive Lyapunov exponent along a given Lagrangian trajectory gives us the maximal rate of stretching along that trajectory. We should, therefore, expect the bulk magnetic field to be stretched out maximally along $\mathbf{e}_1$ on that trajectory. The growth in magnetic field strength then depends on the rate at which the field is twisted in its transverse direction. As a result, Zeldovich et al (1984) deduced that the growth of magnetic energy takes place as $B^2 \sim \exp[(\lambda_1 - \lambda_2)t/2]$. As a result, $\langle |B| \rangle$ should have a growth rate of $\langle (\lambda_1 - \lambda_2)/4 \rangle$ where the average is taken over the ensemble of particles. (Because it is often the case that $|\lambda_1| > |\lambda_2|$, especially in incompressible flows, some authors have tried to relate the growth of magnetic energy to $\lambda_1$ alone.) We see, therefore, that the first two Lyapunov exponents as well as the first Lyapunov eigenvector play a significant role in our understanding of magnetic field amplification in the small scale dynamo.

The Lyapunov dimension has also been defined in Ott (1993) and gives us a good measure of the dimensionality of the stretching that takes place in turbulent flows. For any given Lagrangian test particle, let M be the largest integer such that

$$\sum_{j=1}^{M} \lambda_j \geq 0$$

The Lyapunov dimension for the flow in which that particle is carried is given by

$$D_L = M + \frac{1}{|\lambda_{M+1}|} \sum_{j=1}^{M} \lambda_j$$

For incompressible flows, with their constraint that the sum of the Lyapunov exponents is zero, the Lyapunov dimension can only be 3. For compressible flows it can range between 0 and 3, highlighting one of the fundamental points of difference between compressible and incompressible flows. In the paragraphs below we apply the ideas built up in this and the previous paragraphs to the dynamo fields that are generated in our simulation.

To make a practical evaluation of the Lyapunov exponents, we initialize a swarm of $64^3$ particles so that we have sufficient statistics with which to construct PDFs. Since these particles are initially laid out on a rectangular mesh, the evaluation of Lyapunov exponents shows an initial imprint of the coherency of neighboring eddies. However, as was shown in Section III, after a time interval of 0.6 Myr that coherency between the turbulent eddies begins to fade away. For that reason we initialized the particles at a time of 20.0 Myr after the start of the simulation and displayed the cumulative Lyapunov exponents at a time of 20.6 Myr, i.e. we use an interval over which some coherency is preserved. Figs. 9a to 9d show the PDFs of each of the three cumulative Lyapunov exponents, $\lambda_1, \lambda_2, \lambda_3$, and $(\lambda_1 - \lambda_2)/2$ respectively at a time of 20.6 Myr in the simulation. We see that a significant number of particles have $\lambda_1 > 0$, indicating that we have a substantial amount of field line stretching in the flows generated by the small scale dynamo. We also see that the PDF for $\lambda_1$ in Fig. 9a shows a broad tail, especially at high values of $\lambda_1$. This is significant because it shows us that several portions of magnetic field undergo stretching that is substantially greater than the mean amount of stretching that can be discerned from Fig. 9a. As a result, magnetic fields that are on rapidly stretching trajectories get strongly amplified to magnitudes that are substantially larger than the mean magnetic field, thus accounting for the long tails that are seen in the PDFs for $|\mathbf{B}|$ in Fig. 4. These locally concentrated magnetic fields have a small volume filling factor and, therefore, account for the high intermittencies that we observed for the magnetic field in Fig. 3a. The PDFs in Figs. 9b and 9c also show broad tails showing that in high Mach number flows the contributions from the other two Lyapunov exponents cannot be ignored. From Fig. 9d we also see that for a large fraction of particles we have $\lambda_1 - \lambda_2 > 0$, indicating that the magnetic fields along those trajectories are likely to be amplified. Using the PDF in Fig. 9d we find $\langle (\lambda_1 - \lambda_2)/4 \rangle$ to be given by $(1.9\text{Myr})^{-1}$. The growth rate that we obtain from the Lyapunov exponents agrees to within an order of magnitude with the growth rate of $(20.3\text{Myr})^{-1}$ that we measured in Sub-section II.a for the magnetic flux and the growth rate of $(10\text{Myr})^{-1}$ that we measured for the magnetic energy in Paper I. Our detailed study of Lagrangian chaos has, therefore, given us significant insights into the growth rates for magnetic field in the small scale interstellar dynamo.

For times much smaller than a coherence time we expect the magnetic fields to be stretched in the direction of the first Lyapunov eigenvector. Fig. 10 shows the PDF of $|\mathbf{e}_1 \cdot \mathbf{b}|$ with $\mathbf{b} = \mathbf{B}/|\mathbf{B}|$ which gives the distribution of angles that the magnetic field makes with the dominant direction of stretching. Here $\mathbf{e}_1$ was taken to be the first cumulative Lyapunov eigenvector evaluated by using our ensemble of particles at a time of 20.05 Myr. Fig. 10 shows that the magnetic fields are indeed aligned with the first Lyapunov eigenvector because the PDF of $|\mathbf{e}_1 \cdot \mathbf{b}|$ shows a substantial concentration in the vicinity of $|\mathbf{e}_1 \cdot \mathbf{b}| = 1$. A random orientation of magnetic field would have produced a flat PDF and Fig. 10 shows that such is not the case. We see, therefore, that our detailed study of Lagrangian chaos also enables us to understand the structure of the magnetic field. The magnetic field seeks to align itself with the first eigenvector of the displacement gradient matrix.

It is also interesting to analyze the dimensionality of the turbulence. Fig. 11 shows the PDF of the Lyapunov dimension number at a time of 20.6Myr in the simulation. We see that there is a range of dimension numbers, though most of the dimension numbers are either 3 or 2. To help the reader develop an intuitive understanding of the concept of Lyapunov dimension number, we make a sequence of inexact statements which, nevertheless, vividly illustrate the concept of a dimension number. Situations where all three Lyapunov exponents are positive correspond to flows that expand in all three directions corresponding to a Lyapunov dimension of 3. Such flows dilute out the field. Flows with two positive Lyapunov exponents and one large negative exponent are likely to have a Lyapunov dimension that is close to 2 and would concentrate magnetic fields into sheet-like structures. Flows with one positive Lyapunov exponent and two large negative ones are likely to have a Lyapunov dimension of 1 and would concentrate magnetic fields into narrow ropes. Flows with all negative Lyapunov exponents correspond to a Lyapunov dimension of 0 and correspond to regions of compression in all directions. The previous statements about the Lyapunov dimension number were illustrative and we can see from its definition that the dimension number really depends on the signs and relative magnitudes of all the Lyapunov exponents. By virtue of its definition, the Lyapunov dimension number for a compressible flow can assume all real numbers between 1 and 3 and, because of a peculiarity in the definition, a Lyapunov dimension number of 0 is also possible. In contrast, the Lyapunov exponents in an incompressible flow sum to zero; thus an incompressible flow always has a fixed Lyapunov dimension number of 3. While incompressible flows can also compress magnetic fields into sheets and ropes, the incompressibility condition limits the sum of the Lyapunov exponents in the stretched dimension (dimensions) to always equal the absolute value of the sum of the Lyapunov exponents in the contracted dimensions (dimension). This limits the amount of field line stretching that can take place in an incompressible flow. There are no such bounds for compressible flows. We see from Fig. 11 that the full gamut of Lyapunov dimensions are achieved, indicating that a range of magnetic structures are formed by the compressible, small scale dynamo. While a majority of the particles have a dimension number of 3, we see that a dimension number of 2 is also frequently achieved, indicating that SN-driven turbulence also has a strong tendency to form sheet-like structures in the magnetic field. This was demonstrated visually in Fig. 11 of Paper I and the present section allows us to quantify that fact. We, therefore, realize that the magnetic field growth in our simulations takes place in a flow that is truly topologically different from incompressible flow.

**IV.c) Cancellation Exponent**

In Sections IV.a and IV.b we saw that the line stretching and the Lyapunov exponents overestimate the magnetic field growth. The reason is that they both measure the maximal rate of stretching. But to get growth of magnetic field, we also need constructive folding in the flows that sustain a dynamo. The STF dynamo from Fig. 1.3 of CG is one example of a dynamo with perfectly constructive folding. Most fluid dynamical motions do not result in such perfectly constructive folding and many of the motions in interstellar turbulence destroy magnetic field almost as much as they help create it. It is for this reason that the field growth in the small scale dynamos presented here is much smaller than the line stretching exponent or the Lyapunov exponents. The cancellation exponent, $\kappa$, measures the amount of small scale field structure that does not contribute to a large scale flux and, therefore, measures the efficiency or inefficiency of constructive folding, see Ott et al (1992) and Du and Ott (1993a, b). To get the cancellation exponent associated with a certain magnetic field $\mathbf{B}(\mathbf{x})$ we define its convolution $H_{l^2}\mathbf{B}$ with the kernel of the diffusion operator with a diffusion length scale "$l$" as follows:

$$H_{l^2}\mathbf{B}(\mathbf{x}) = \frac{1}{l\sqrt{2\pi}} \int_V \exp[-(\mathbf{x}-\mathbf{y})^2/(2l^2)]\, \mathbf{B}(\mathbf{y})\, d\mathbf{y}$$

i.e. the convolution smoothes out magnetic structures on the length scale "$l$". The magnetic field structure on lengths larger than "$l$" is measured by:

$$\xi(V,l) = \frac{\int_V \left|H_{l^2}\mathbf{B}(\mathbf{x})\right| d\mathbf{x}}{\sqrt{\left|\int_V H_{l^2}\mathbf{B}(\mathbf{x})\, d\mathbf{x}\right|}}$$

If the dynamo field has a significant amount of small scale structure, due perhaps to a degradation in constructive folding, we expect $\xi(V,l)$ to have as strong variation with "$l$", with smaller values of "$l$" producing substantially larger values of $\xi(V,l)$ than larger values of "$l$". On the other hand, in the limit of a dynamo with perfectly constructive folding (which is almost never achieved for real flows) we expect $\xi(V,l)$ to show almost no *l*-dependence. The cancellation exponent κ is given by the scaling behavior :

$$\xi(V,l) \sim P\, l^{-\kappa} \quad \text{or} \quad \kappa = \frac{\log(\xi(V,l))}{\log(1/l)}$$

where P is a constant. It might now seem intuitively evident that the growth rate of magnetic field in the dynamo might in some way be governed by the largest Lyapunov exponent $\lambda_1$ while the destruction of field by less-than-perfect constructive folding is governed by some combination of the smallest Lyapunov exponent $\lambda_3$ and the cancellation exponent κ . It would also seem reasonable that the mean dynamo growth rate should be considered over an ensemble average of Lagrangian trajectories. Du and Ott (1993b) considered dynamo growth rates for incompressible flows. Corresponding results for compressible flows are not available. Thus we have no option but to carry over the incompressible result from Du and Ott (1993b) even though its strict applicability to compressible flows is not justified. The dynamo growth rate that factors in constructive folding, $\gamma_{0,\,CF}$, is related to an ensemble average over the particles we considered in Section IV.b and is given by:

$$\begin{aligned}\gamma_{0,\,CF} &= \lim_{T\to\infty} \frac{1}{T} \log \left\langle \exp[\lambda_1 T] \exp[\kappa\, \lambda_3\, T] \right\rangle \\ &\approx \frac{1}{\tau_{coherence}} \log \left\langle \exp[\lambda_1\, \tau_{coherence}] \exp[\kappa\, \lambda_3\, \tau_{coherence}] \right\rangle\end{aligned}$$

Since taking the limit with $T \to \infty$ is unwieldy in the first of the two equations above, we approximate the limit by considering a coherence time, $\tau_{coherence} = 0.6\text{Myr}$ , in the flow.

We applied the above techniques to the same swarm of $64^3$ particles that we mentioned in Section IV.b. Fig. 12 shows the variation of $\xi(V,l)$ v/s "*l*" at a time of 20.6 Myr. Using Fig. 12 we found the cancellation exponent in the magnetic field to be κ = 0.8

. We see, therefore, that the realistic small scale dynamo that we are exploring here shows very poor constructive folding. This is in keeping with expectation. On carrying out the ensemble averages for the growth rate, we find a growth rate of $\gamma_{0,\,CF} = (0.1\,\mathrm{Myr})^{-1}$ which does not compare favorably with the growth rate for the magnetic flux that we found in Section II.a and the growth rate for the magnetic energy that we found in Paper II. Just like the line stretching exponent, $\gamma_{0,\,CF}$ does, however, provide a good upper bound on the growth rates for magnetic field and magnetic flux. The lack of good agreement is because the above formula for $\gamma_{0,\,CF}$ is strongly dependent on assuming an incompressible flow. The cancellation exponent is, nevertheless, a very important way of measuring the structure of the magnetic field and relating it to dynamical chaos. The equations for κ given in the previous paragraph can be directly translated by observers to obtain an observed estimate of the cancellation exponent in our Galactic magnetic field.

## V) Discussion and Conclusions

### V.a) Discussion

We have shown in Paper I that small scale magnetic fields can be made to grow very rapidly in our Galactic ISM. Observations of high redshift galaxies also show that star formation takes place very rapidly in those environments, see Churchill & Le Brun (1998) and Ledoux, Srianand & Petitjean (2002). A substantial part of the difficulty in forming low mass stars has to do with the removal of angular momentum from the protostellar cores. Paleologou & Mouschovias (1983) have shown that magnetic fields can help remove the angular momentum from such cores. Thus if a SN-driven fast dynamo could operate in protogalactic environments, it could rapidly form magnetic fields thus enhancing the formation of low mass stars. The fields that are formed in our simulations have length scales as large as a hundred parsecs, which are certainly larger than the characteristic 0.1 pc size of a protostellar core. Thus the small scale dynamo explored here can play an important role in star formation in out proto-Galaxy and also in high redshift galaxies. The length scales over which magnetic fields form are also comparable to the size of a globular cluster or a giant molecular cloud indicating that such magnetic fields could also have contributed to low mass star formation during the early phase of Galaxy formation.

Nakano, Nishi & Umebayashi (2002) have also shown the importance of ambipolar drift in removing magnetic fields in later stages of low mass star formation. The availability of metals and possibly grains plays an important role in this process. SN ejecta play an important role in providing that metallicity. Efficient mixing of SN ejecta by way of SN-driven turbulence is important in ensuring that the metals are well-mixed, thus leading to efficient star formation. A quantitative measure of turbulent diffusion enables us to quantify the rate of turbulent mixing. We have, therefore, come up with a strategy for directly reading off the turbulent diffusivity from numerical simulations. Our strategy can be applied to a large number of interstellar simulations, enabling us to obtain the dependence of the turbulent diffusivity of the ISM on its SNe rate and its mean density, pressure and magnetic field. A similar exercise can be carried out for proto-Galactic environments.

The variation of the magnetic field strength with density that we find in our simulations goes as $|B| \sim \rho^\alpha$ with $\alpha = 0.386$. This is quite similar to the variation found by Crutcher (1999) in molecular clouds where $|B| \sim \rho^\alpha$ with $\alpha = 0.47$. Crutcher (1999) found equipartition magnetic fields that range from tens to thousands of µG in the clouds. The processes that drive a mean field dynamo in our Galaxy, i.e. the mean helicity from buoyant supernova and superbubble shells and the mean shear from Galactic differential rotation, are not available in a molecular cloud. Yet, the data shows that magnetic fields in molecular clouds are in equipartition with the turbulence. Since the turbulence in molecular clouds is known to be supersonic and proto-stellar jets and winds can drive shocks in such a medium, it is very likely that magnetic fields can be amplified in molecular clouds via the small scale dynamo processes described here. The fact that the strong magnetic fields in the small scale dynamo concentrate themselves in the higher density gas in the current simulations would become even more interesting if the trend extended to the cold molecular gas. The cooling curve in the present simulations does not make it possible for such gas to form, however, if the trend should continue to the cold phase then it would give credence to the suggestion in Kulsrud & Anderson (1992) that the cold phase might play an important role in generating the magnetic field of the Galaxy.

In their study of the magnetic structure of the turbulence in our Galaxy, Minter & Spangler (1996) find that there is a critical length scale, which they find to be 4 pc, below which the turbulence begins to become three-dimensional. Fosalba et al (2002) suggest that the turbulence is three-dimensional on even larger scales. The ability to measure $\frac{\langle B^2 \rangle}{\langle \mathbf{B} \rangle^2}$ directly from simulations and to relate that to the ratio of synchrotron emission measure to the pulsar rotation measure is very interesting. It enables a direct contact between observations and simulations. Likewise, it would be very interesting to try and obtain observational measures of the kurtosis $\frac{\langle B^4 \rangle^{\frac{1}{2}}}{\langle B^2 \rangle}$ and relate that to the values found from simulations. Such observations would provide fresh insights into the intermittency of the interstellar turbulence. The PDFs of density and magnetic field can also be compared with observations, providing further insights. Observational demonstration of the existence of a high-end tail in the PDF of the magnetic field strength while simultaneously demonstrating that the density PDF shows no such tail would make for a dramatic demonstration that a small scale turbulent dynamo might be operating in our Galactic ISM. In Paper I we have already provided spectra for the magnetic energy and the density fluctuations, further facilitating direct comparison with observations such as Minter & Spangler (1996) and Armstrong, Rickett & Spangler (1995). It is even possible to find observational analogues to our technique for measuring the cancellation exponent. Such observations would yield a direct link between observations and the chaotic structure of flow lines within our Galactic ISM.

Studies of open clusters, such as the Pleiades cluster by Wilden et al (2002), have shown that the resulting stars have nearly identical metallicities. Observations of low metallicity halo stars by Arnone et al (2005) also seem to point to a similar conclusion. If star formation takes place in a turbulent, molecular cloud-like environment then our study of line stretching shows that the diffusion of elements down to the molecular level takes place very efficiently in such environments, lending support to rapid and efficient mixing of elements in the progenitor gas clouds for those stellar populations. The characteristic time scales that we find in sub-section IV.a are much smaller than the anticipated times for the formation of these systems. The dependence of this mixing time on ISM parameters and SNe rates is a topic of future study. However we can say that that if star formation takes place on timescales that are much larger than a few times 0.20Myr in a system quite like the one that we have simulated then we can conclude that the environment in which that star formation takes place is likely to be chemically well-mixed.

The formation of thin, narrow structures and the elevated turbulent diffusion coefficient, as presented in Sections III and IV.a, are also relevant to heat transport in the ISM. Heat transport by thermal conduction is important because it regulates the formation of mixing layers as shown by Borkowski, Balbus, & Fristrom (1990) and Begelman & McKee (1990). The temperature of the gas in such mixing layers is intermediate between that of the hot and warm phases. Such gas is observable via high ionization stage ions such as OIV and UV satellites such as FUSE routinely observe such gas, see Heckman et al. (2002). The recent literature on thermal conduction in magnetized plasmas seems to present a divergence of viewpoints. Chandran & Cowley (1998) followed electron trajectories in a single snapshot of tangled, turbulent field lines to claim that the tangling of turbulent field lines diminishes thermal conduction. Cho et al (2003) studied passive scalar diffusion to claim that the transport of gas with high thermal energy by turbulent eddies enhances thermal conduction. The former approach is inadequate because it ignores advection by turbulent eddies. The latter approach is inadequate because it ignores the effect of the parabolic diffusion operator which can even become anisotropic in the presence of magnetic fields. Our simulations here can be used to estimate the high turbulent diffusivities that one might expect in a turbulent ISM, thus lending a measure of plausibility to some of the arguments in Cho et al (2003). The high rates of line stretching would also suggest that thin, narrow regions with alternating high and low levels of thermal energy are likely to form in a turbulent, SN-driven ISM, making it possible for the thermal conduction operator to act on small scales to remove thermal fluctuations. A full resolution of these issues will have to await the incorporation of methods like the ones developed in Balsara & Fisker (2005) and Balsara (2005) into models for the ISM.

**V.b) Conclusions**

We have shown that:

1) The small scale dynamo results in rapid flux growth in times that are comparable to the eddy turnover times in the ISM.

2) The kurtosis is measured from simulations and shows the importance of intermittency in field amplification. It would be interesting to measure the kurtosis via observations of the Galactic ISM.

3) The ratio of the synchrotron emission to the rotation measure can be directly related to diagnostics that can be obtained via our simulations.

4) The density shows a bimodal PDF, consistent with the limitations of the cooling function. The PDF of the magnetic field shows a high-end tail, providing a tell-tale signature of the operation of the small scale dynamo. The magnetic field strength correlates positively with density as $|B| \sim \rho^{\alpha}$ with $\alpha = 0.386$. As a result, the field amplification takes place more vigorously in the lower temperature, denser gas. These findings can be related to observations.

5) The line stretching exponent has been measured and shown to provide an upper bound to the growth rate for magnetic field. The line stretching takes place on timescales that are much shorter than the growth times for the field indicating that the constructive folding in turbulent interstellar flows is very far from perfect.

6) The Lyapunov exponents in turbulent interstellar flows have been measured and their relationship to field growth has been established. The PDF of the Lyapunov dimension number for compressible, interstellar turbulence has been evaluated and it is shown that the turbulence can have a dimensionality that is very different from the incompressible case.

7) The lack of constructive folding has been quantified by way of the cancellation exponent. Use of the cancellation exponent along with the Lyapunov exponents enables us to get a rather precise predictive measure for the growth of magnetic fields in our simulations. In doing so, we establish a strong connection between the Lagrangian chaos that is present in interstellar flows and the growth of the small scale dynamo in such flows. We also point out that the cancellation exponent might be observationally measurable, enabling observations to make an intimate connection with the Lagrangian chaos that we see in our simulations and anticipate in our Galactic ISM.

8) The rate of line stretching is related to the rate at which narrow structures are formed in SN-driven turbulence. The high rates of line stretching in interstellar turbulence suggest that the eventual diffusion of elements at the molecular level is very efficient.

9) We have also obtained a measure of the turbulent diffusivity in our ISM. Its importance in Galactic chemical evolution has also been shown. It is shown that a Lagrangian approach for measuring turbulent diffusivity yields a bounded measure for the turbulent diffusivity. Moreover, the turbulent diffusivity is roughly consistent with the simple mixing length theory developed by Batchelor (1949).

**Acknowledgements** : DSB acknowledges helpful discussions with T. Beers, A. Brandenburg, P. Diamond, D. Galloway, I. Klapper, A. Lazarian, M. Mac Low, M.R.E.Proctor and A. Scheckochihin. DSB acknowledges support via NSF grants R36643-7390002, AST-005569-001 and DMS-0204640. JK was also supported by KOSEF through Astrophysical Research Center for the Structure and Evolution of the Cosmos (ARCSEC). The majority of simulations were performed on PC clusters at UND and KASI but a few initial simulations were also performed at NCSA. The Korean cluster was acquired with funding from KASI and ARCSEC.

**Figure Captions**

Figure 1 shows the the fluxes $\Phi_x$, $\Phi_y$ and $\Phi_z$ through the three $64^2$ zone planes in the center of our computational domain as a function of time.

Figure 2 shows the average growth rate of the maximum of the three fluxes plotted as a function of time. A 5Myr sliding window was used in the averaging process. The time-averaged growth rate is shown by the solid line. The dotted and dashed lines show the time-averaged growth rate in the positive and negative parts of the curve in Figure 2 respectively.

Figure 3a shows the kuritosis in the magnetic field and flow variables, shown via the volume-averaged ratios $\frac{\langle B^4 \rangle^{\frac{1}{2}}}{\langle B^2 \rangle}$, $\frac{\langle B^6 \rangle^{\frac{1}{3}}}{\langle B^2 \rangle}$ and $\frac{\langle v^4 \rangle^{\frac{1}{2}}}{\langle v^2 \rangle}$ as a function of time. Figure 3b shows $\frac{\langle B^2 \rangle}{\langle \mathbf{B} \rangle^2}$ as a function of time. The volume used in the spatial averaging was the $64^3$ zone region that contains the three $64^2$ zone planes from Figure 1 in the center of our computational domain.

Figure 4 shows the PDFs of the absolute magnitude of the magnetic field, $|\mathbf{B}|$, at different epochs in the dynamo simulation. The epochs are 5, 10, 20 and 40 Myr after the start of the simulation.

Figure 5 shows the corresponding PDFs of the density at the same epochs as Figure 4 in the dynamo simulation.

Figure 6 shows the scatter plot of the logarithm of the magnitude of the magnetic field v/s the logarithm of the density at a time that corresponds to the end of the simulation. The hot gas is shown in red and has $\log_{10} T > 5.5$ where "T" is the temperature. The intermediate gas is shown in green and has $3.9 \leq \log_{10} T \leq 5.5$ and is the gas that is not likely to be in thermal equilibrium equilibrium. The warm gas is shown in blue and has $\log_{10} T < 3.9$.

Fig. 7 shows the evolution of the specific turbulent diffusivity, $\eta_{turb}$, as a function of time. Past an eddy coherence time of 0.6 Myr, $\eta_{turb}$ becomes constant.

Fig. 8a shows the evolution of the lengths of line segments along the x,y and z-directions in the flow as a function of time, starting at a time of 20Myr in the simulation. The two thick dashed lines in Fig. 8a mark out line stretching rates of $(0.10\text{Myr})^{-1}$ and $(0.11\text{Myr})^{-1}$. Fig. 8b shows the r.m.s. value of the radii of the points that make up each segment (with the mean subtracted off) as a function of time. Fig. 8c shows the segment that was originally oriented along the y-axis at a time of 20.08Myr, i.e. 0.08Myr after the lines were evolved. Fig. 8d and Fig. 8e show the same segment at times of 20.4Myr and 20.8Myr.

Figures 9a to 9d show the PDFs of each of the three cumulative Lyapunov exponents, $\lambda_1, \lambda_2, \lambda_3$, and $(\lambda_1 - \lambda_2)/2$ respectively at a time of 20.6 Myr in the simulation.

Figure 10 shows the PDF of $|\mathbf{e}_1 \cdot \mathbf{b}|$ with $\mathbf{b} = \mathbf{B}/|\mathbf{B}|$. Here $\mathbf{e}_1$ was taken to be the first cumulative Lyapunov eigenvector evaluated by using our ensemble of particles at a time of 20.05 Myr. We see that there is a strong tendency for the magnetic field vector to be preferentially oriented along the first Lyapunov eigenvector, i.e. the direction along which the stretching is maximal.

Figure 11 shows the PDF of the Lyapunov dimension number at a time of 20.6 Myr in the simulation.

Figure 12 shows the variation of $\xi(V,l)$ v/s "$l$" at a time of 20.6 Myr.

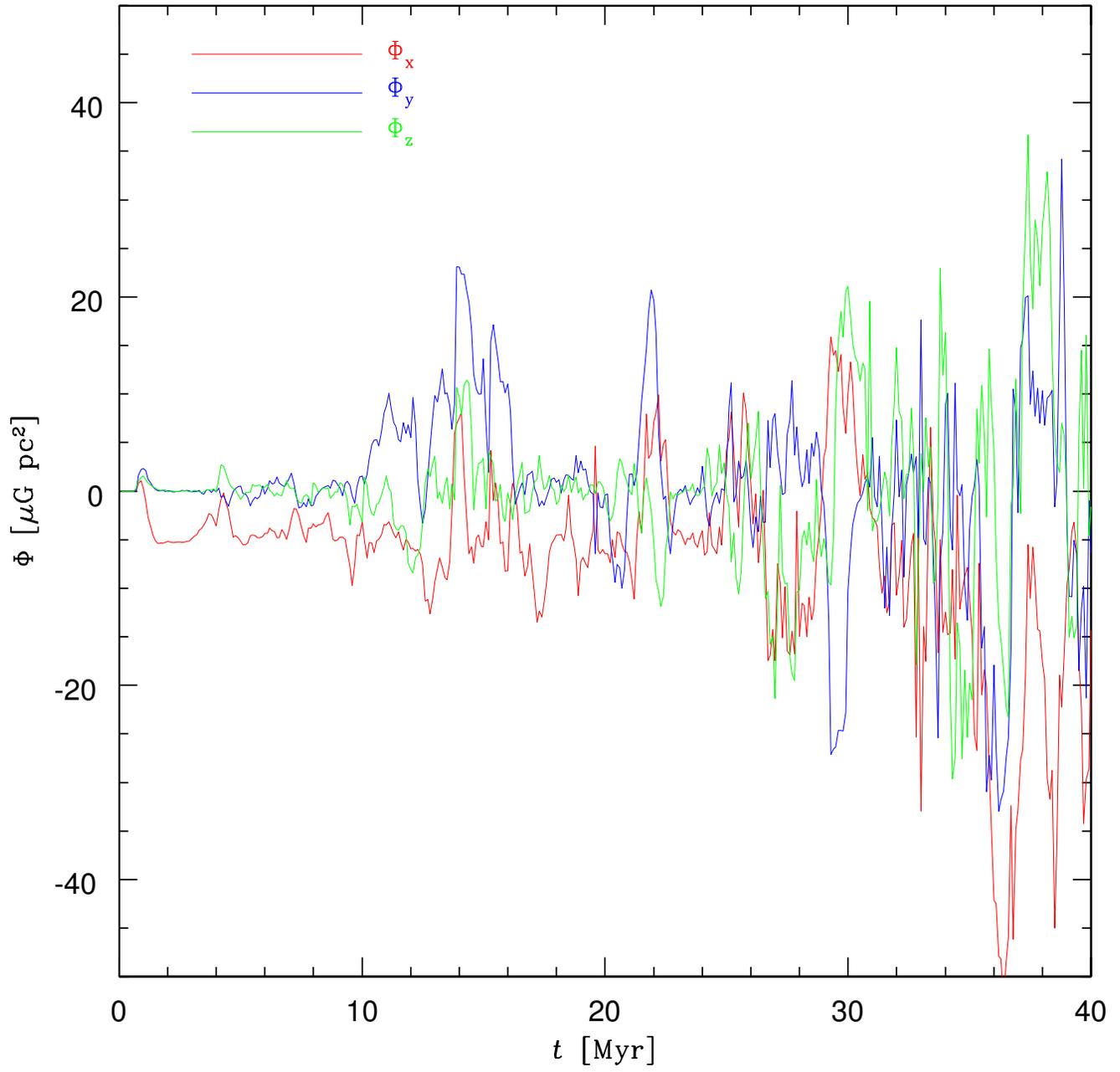

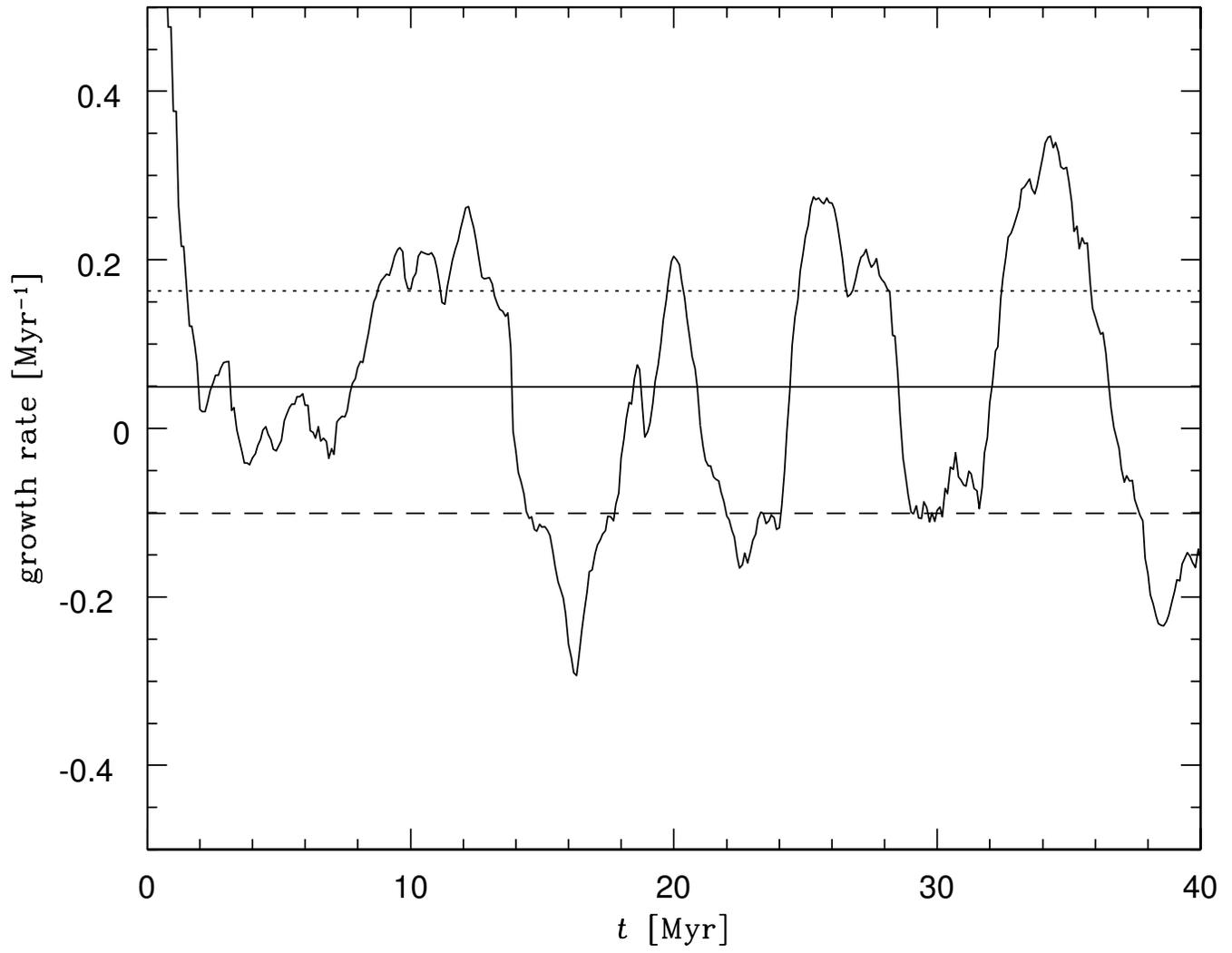

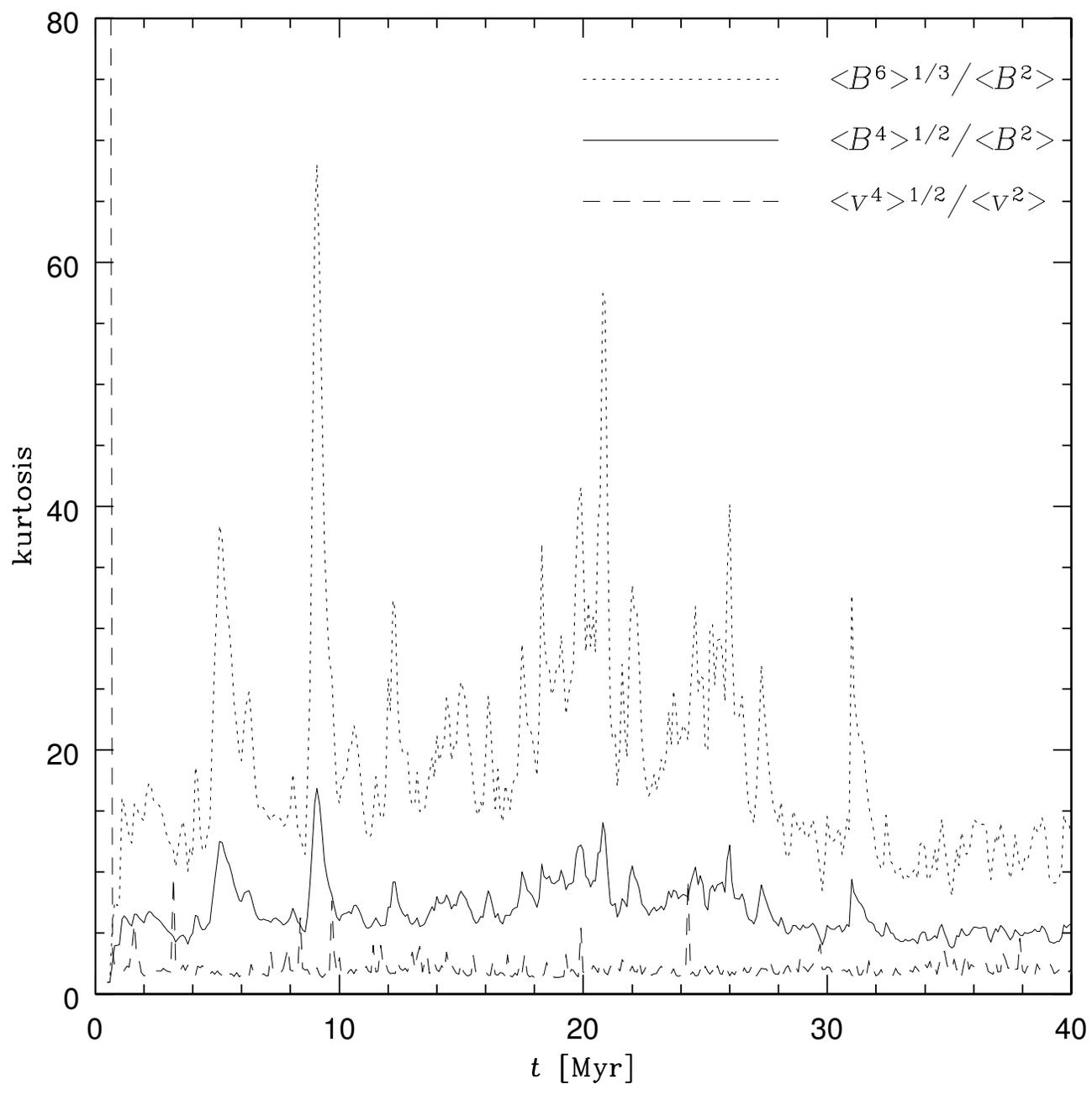

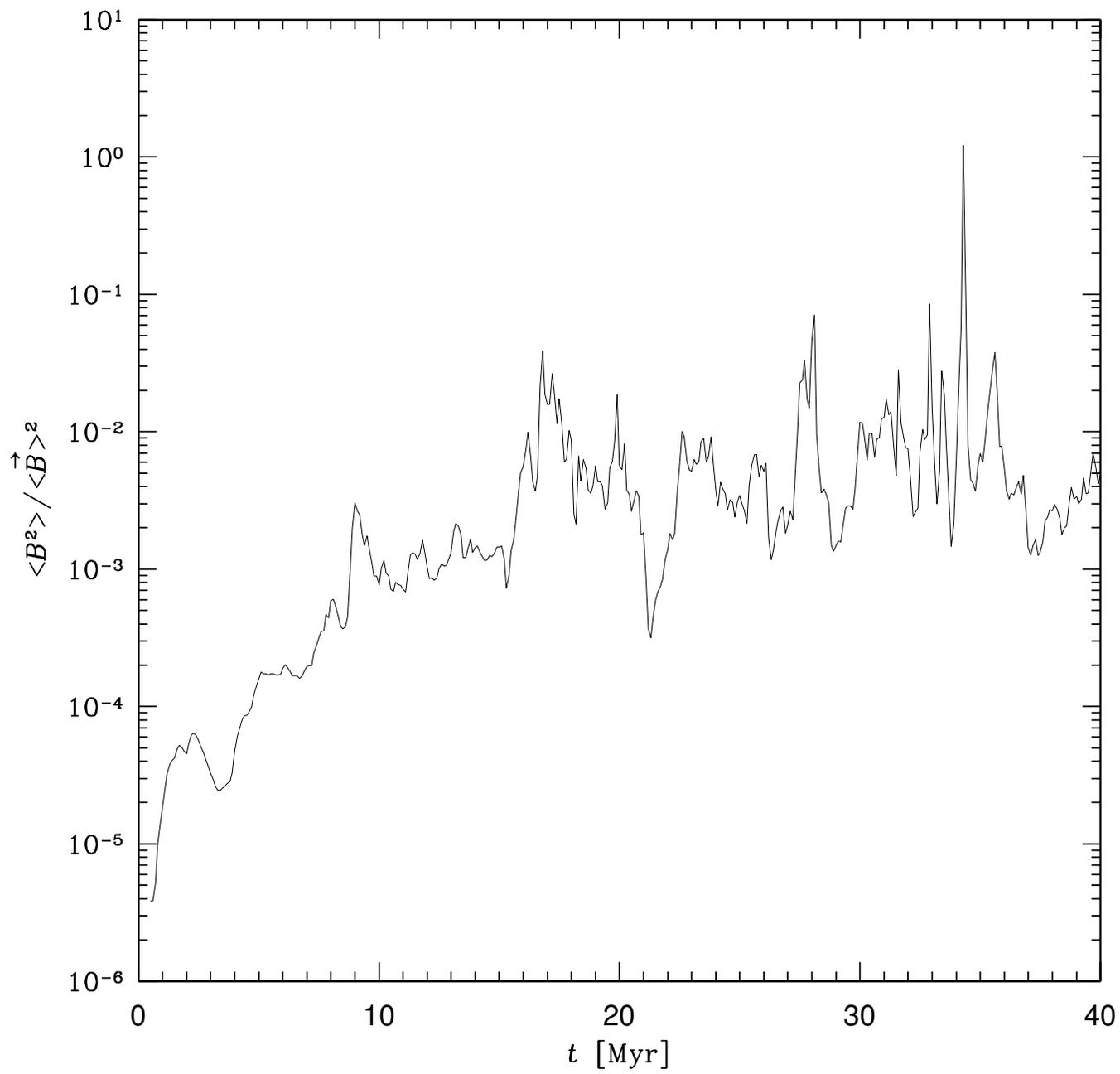

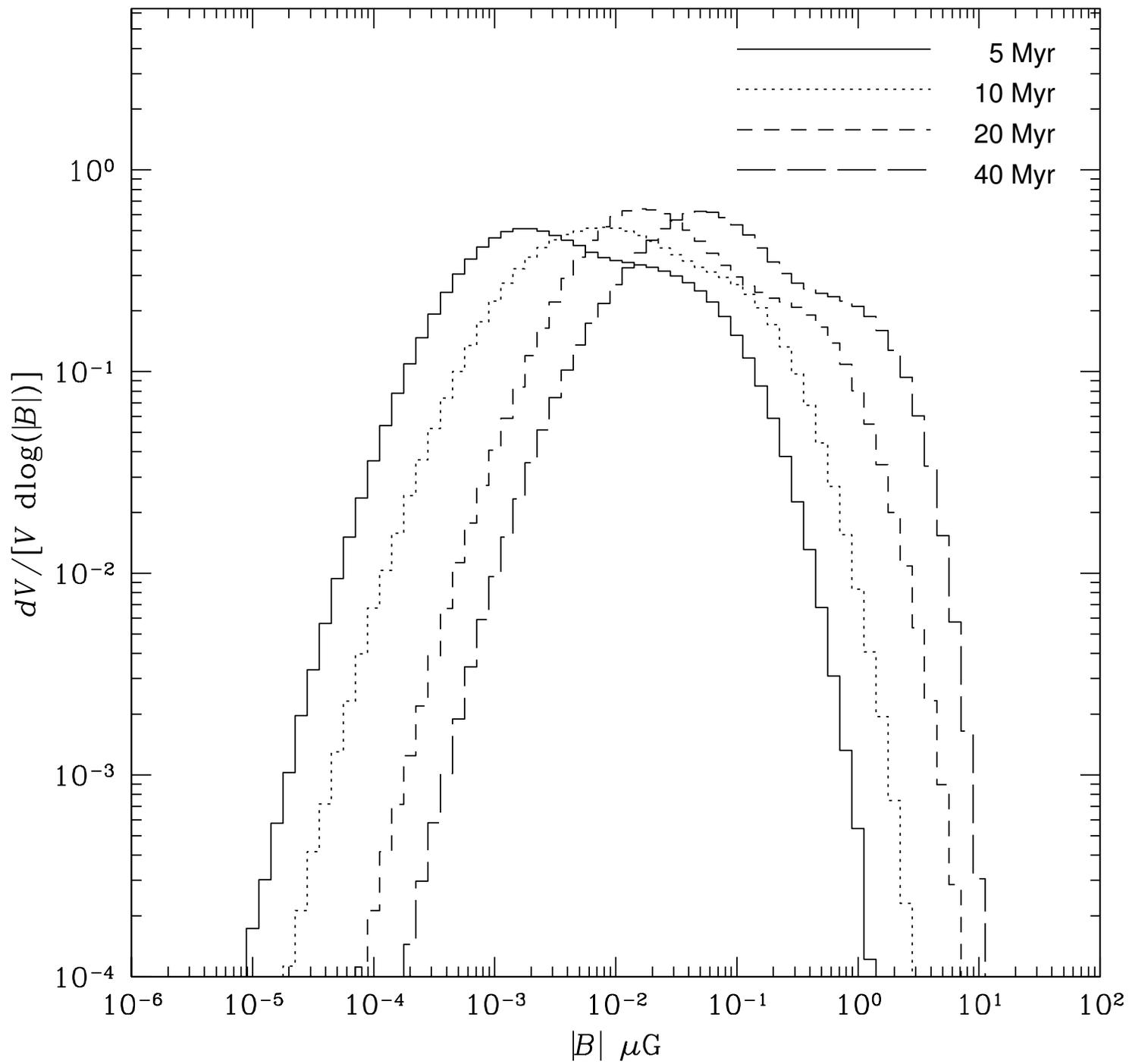

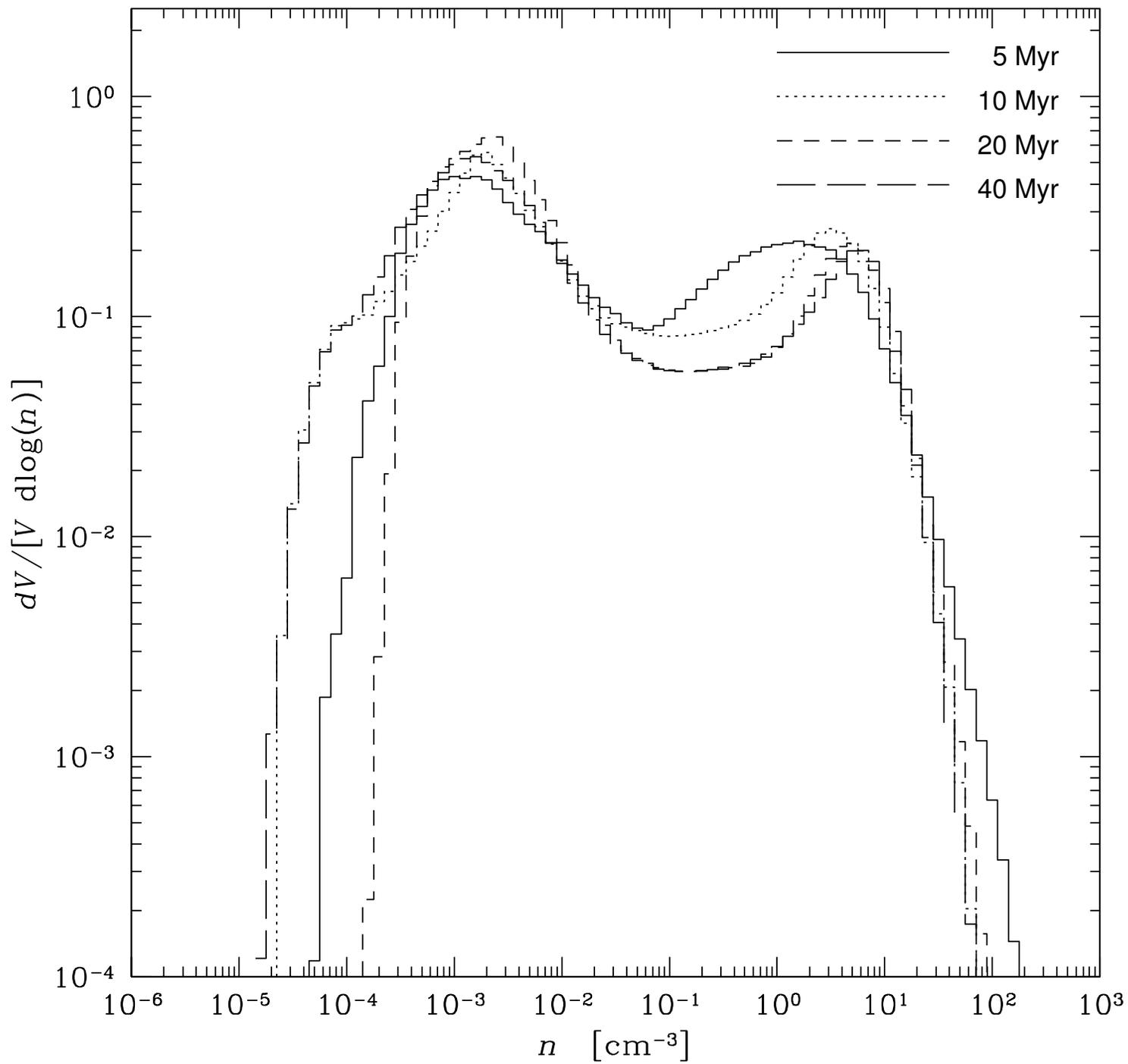

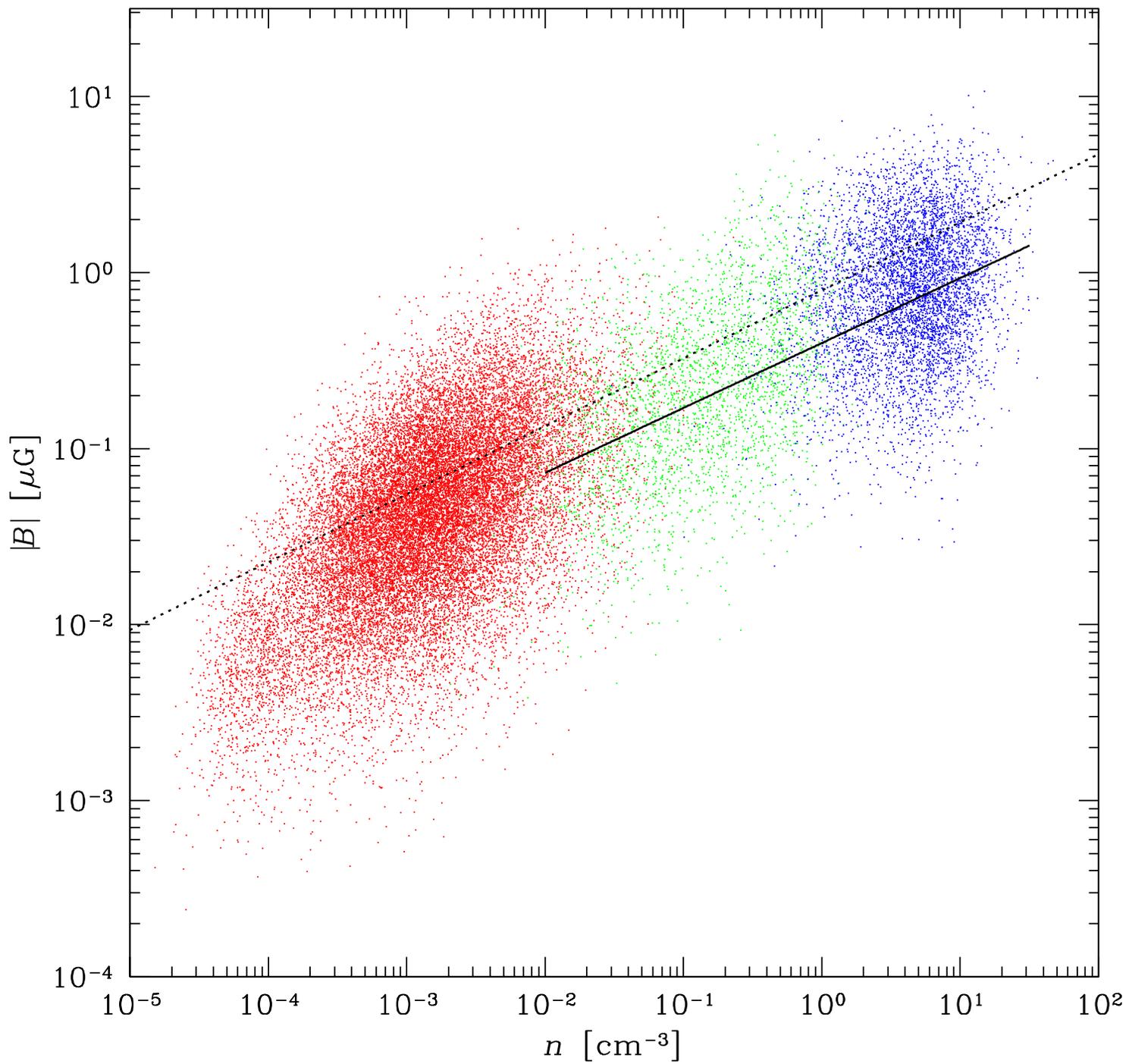

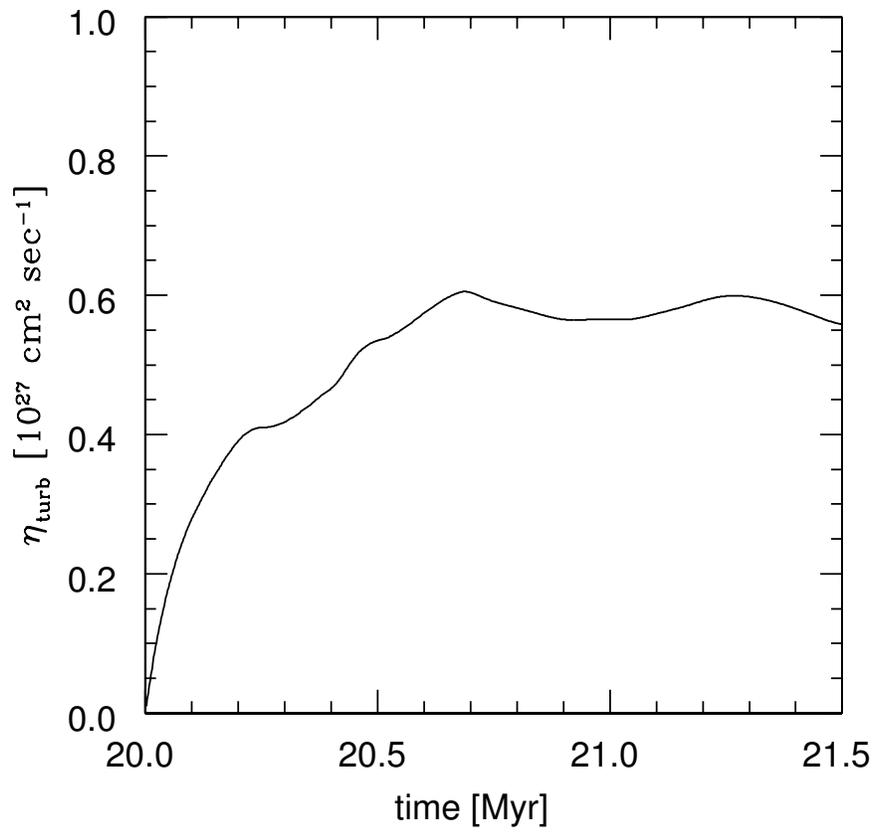

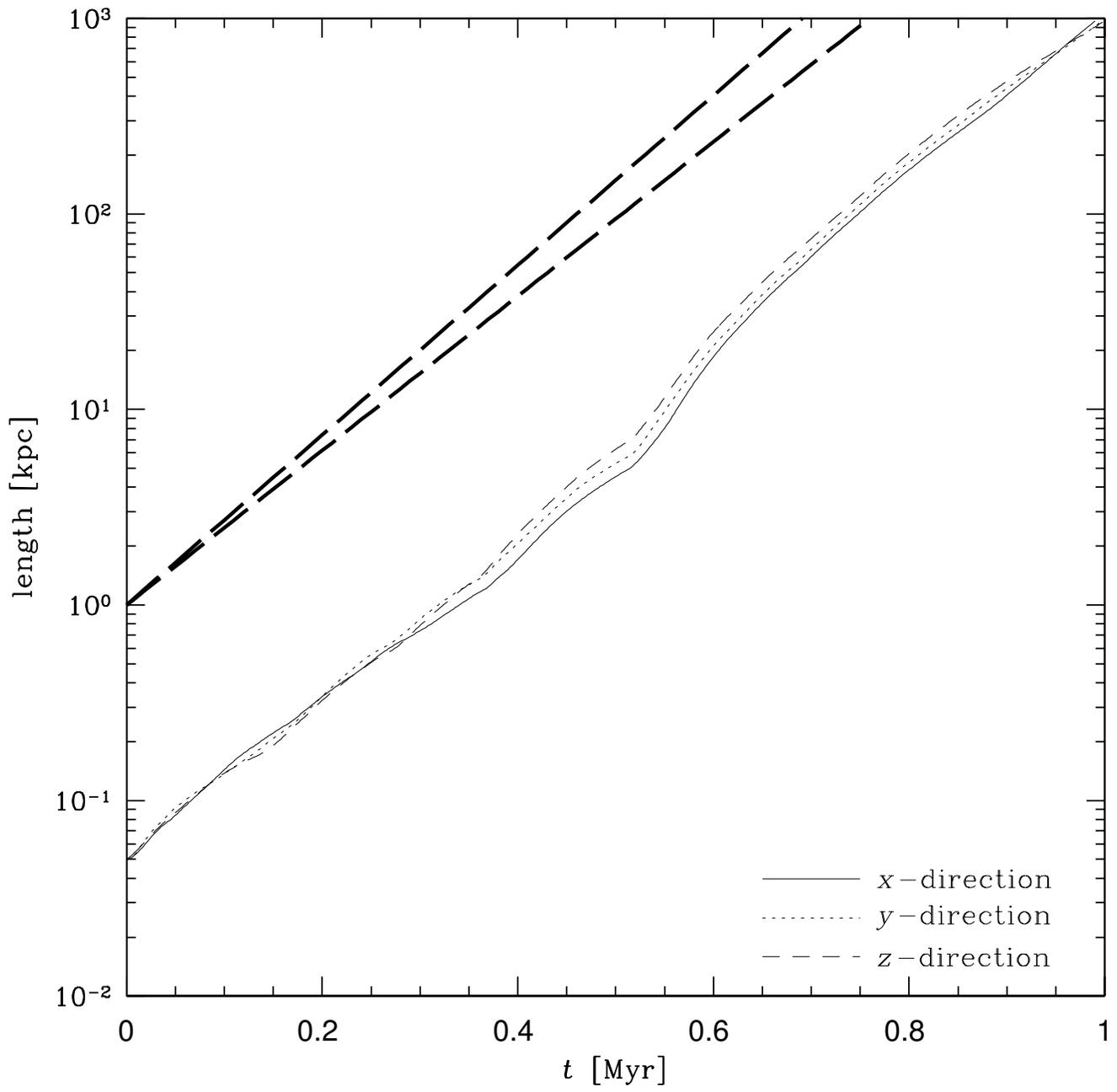

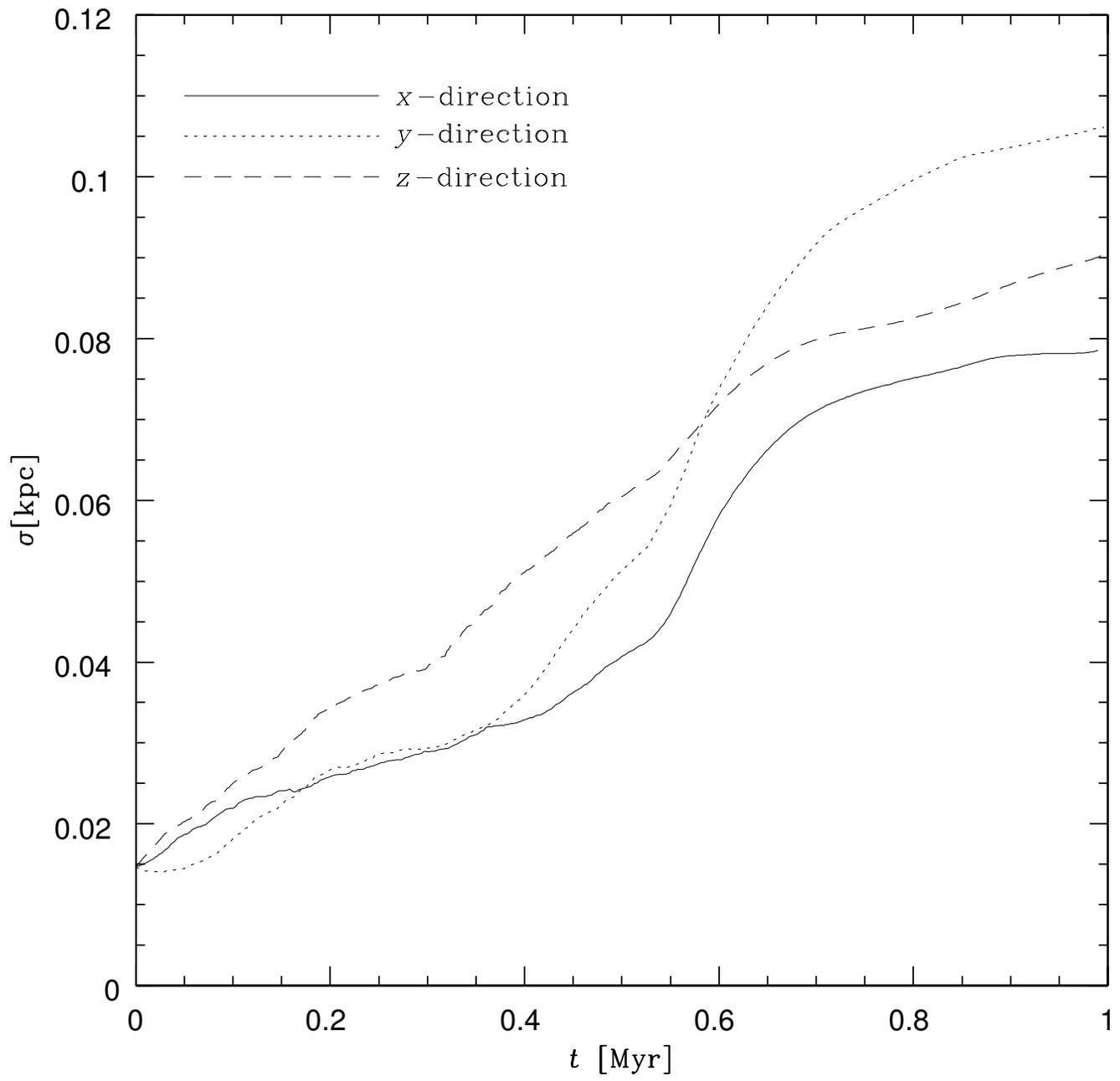

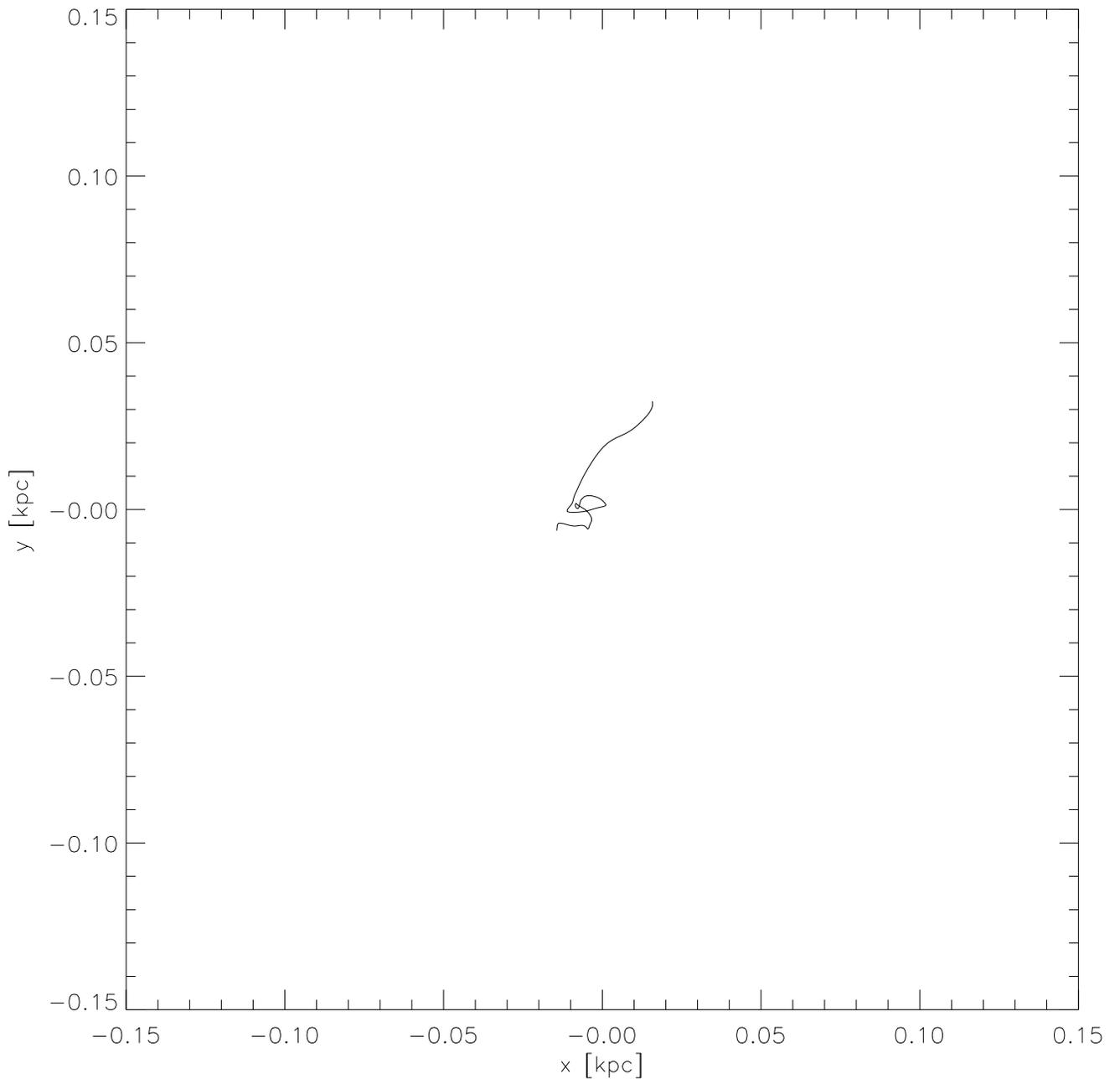

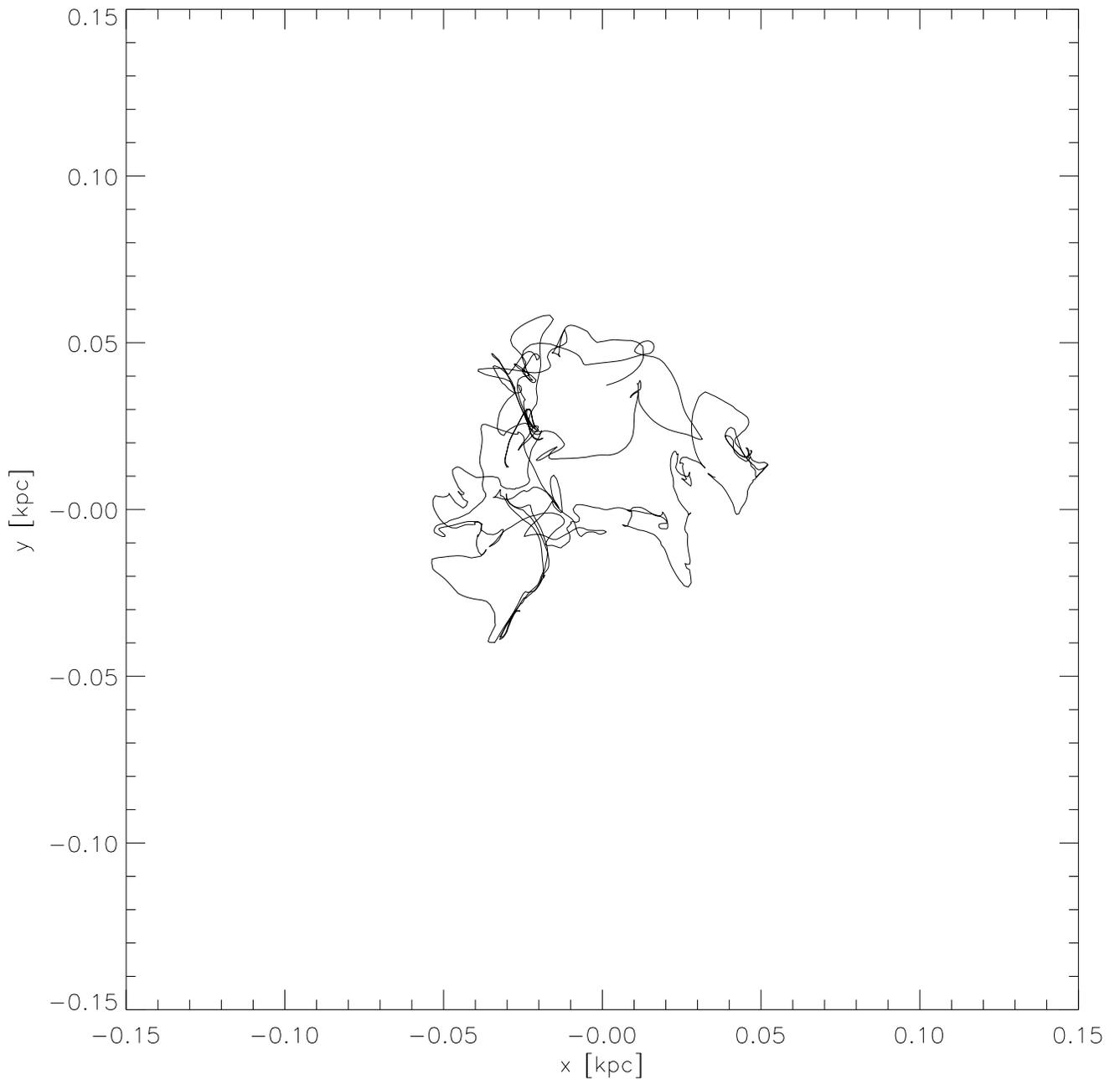

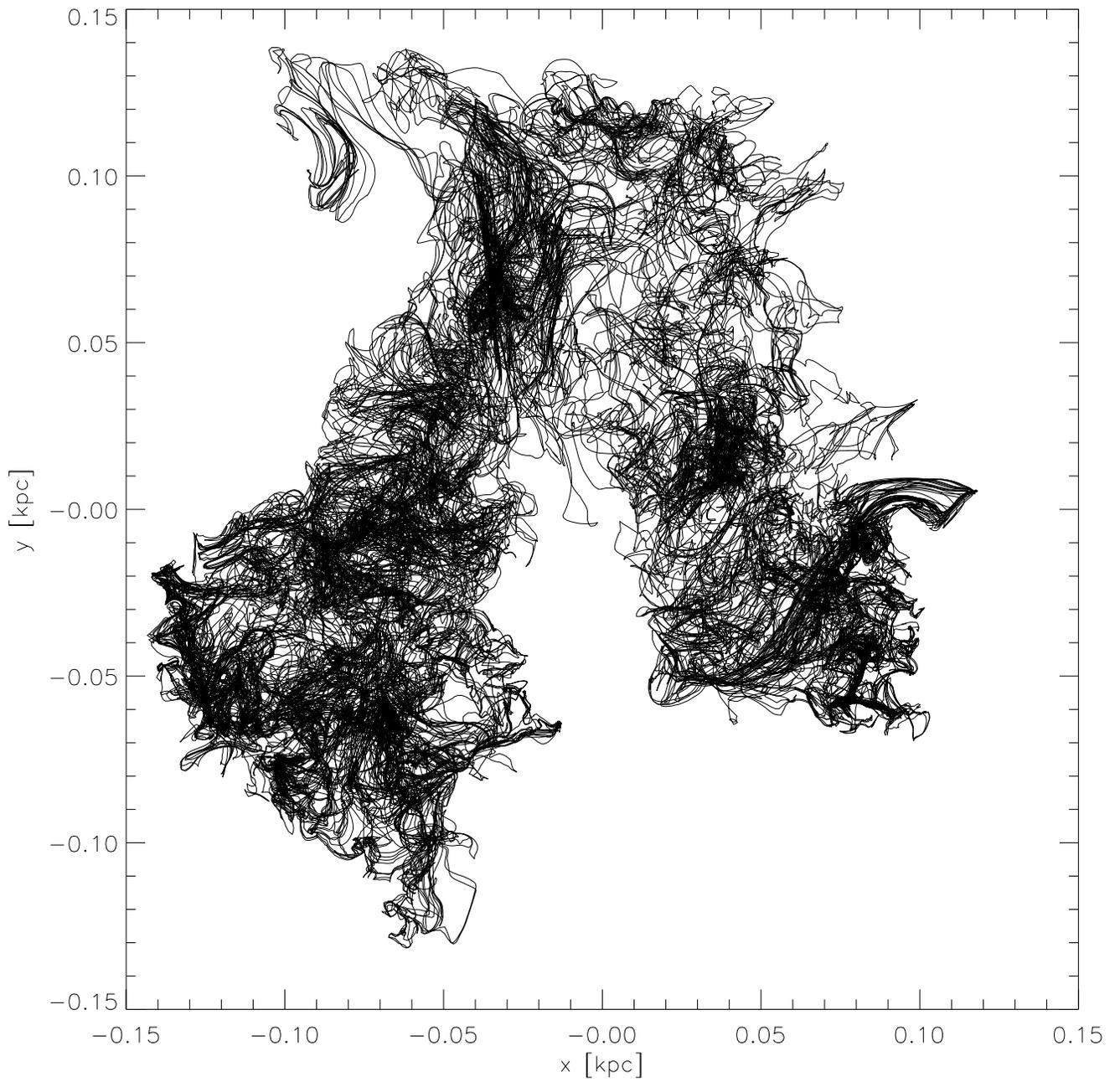

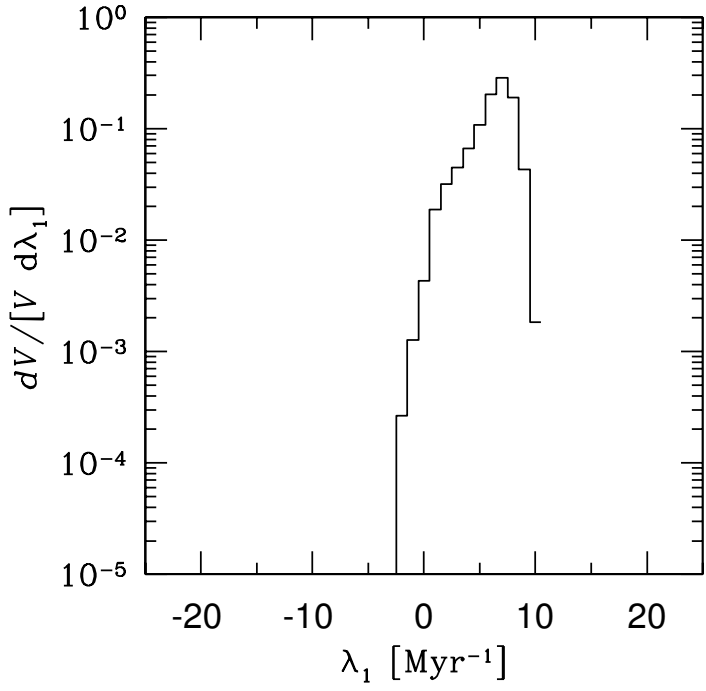
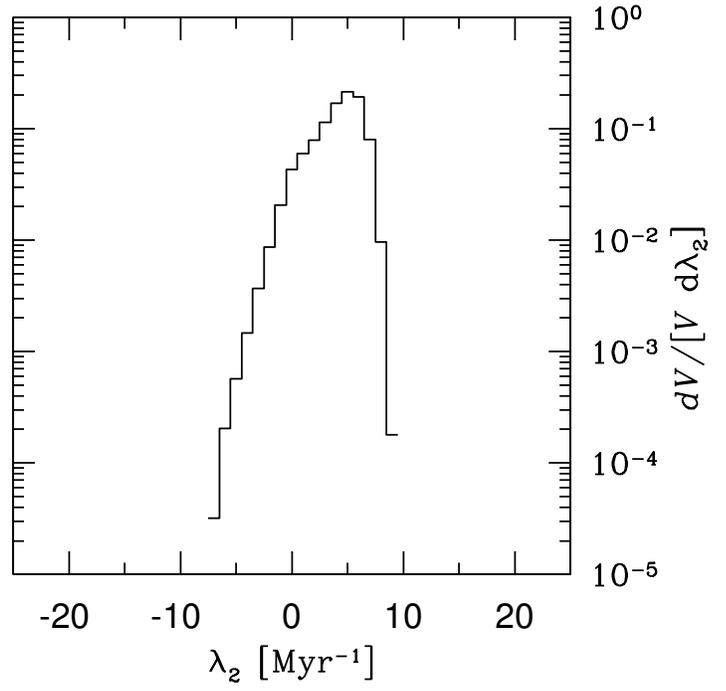
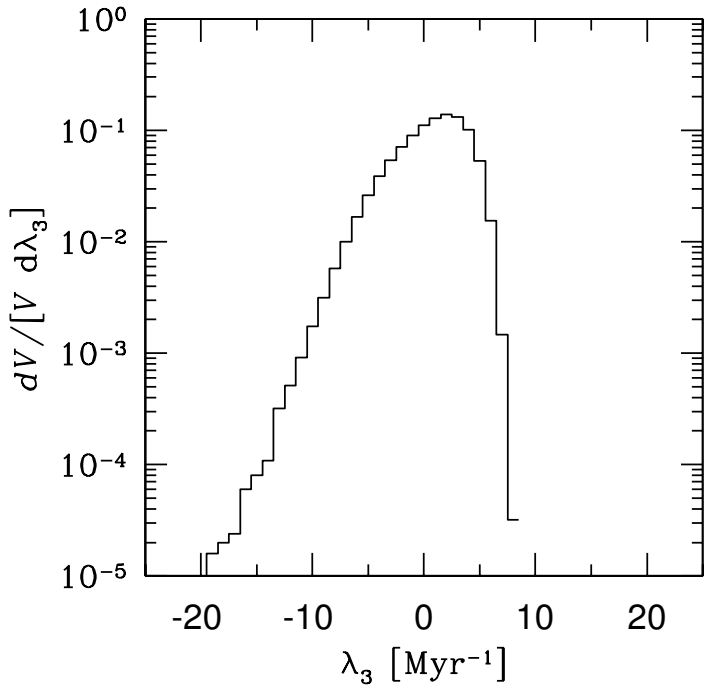
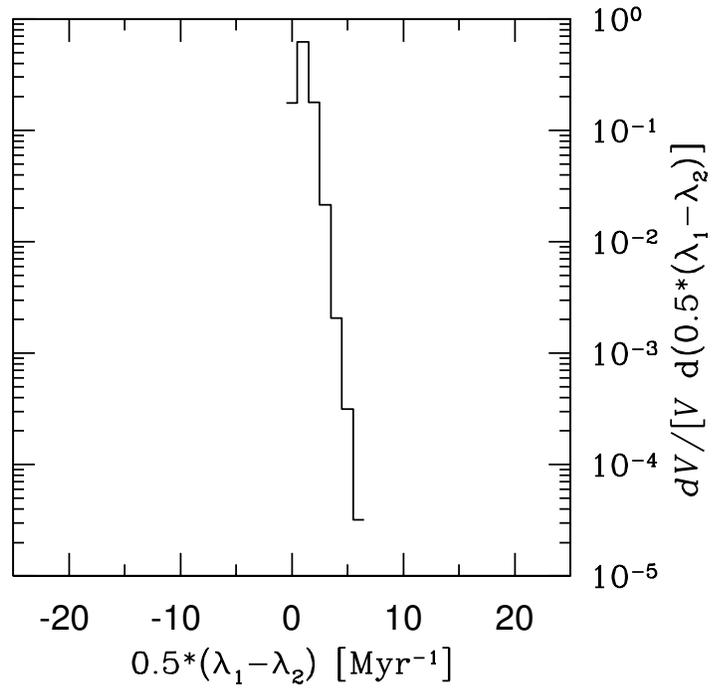

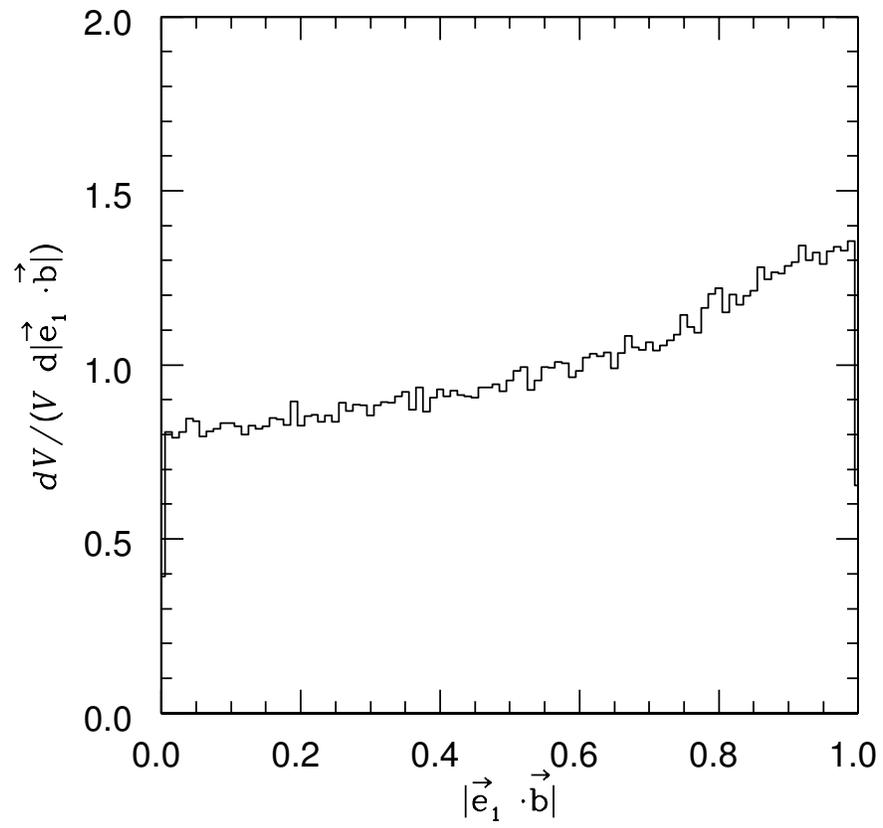

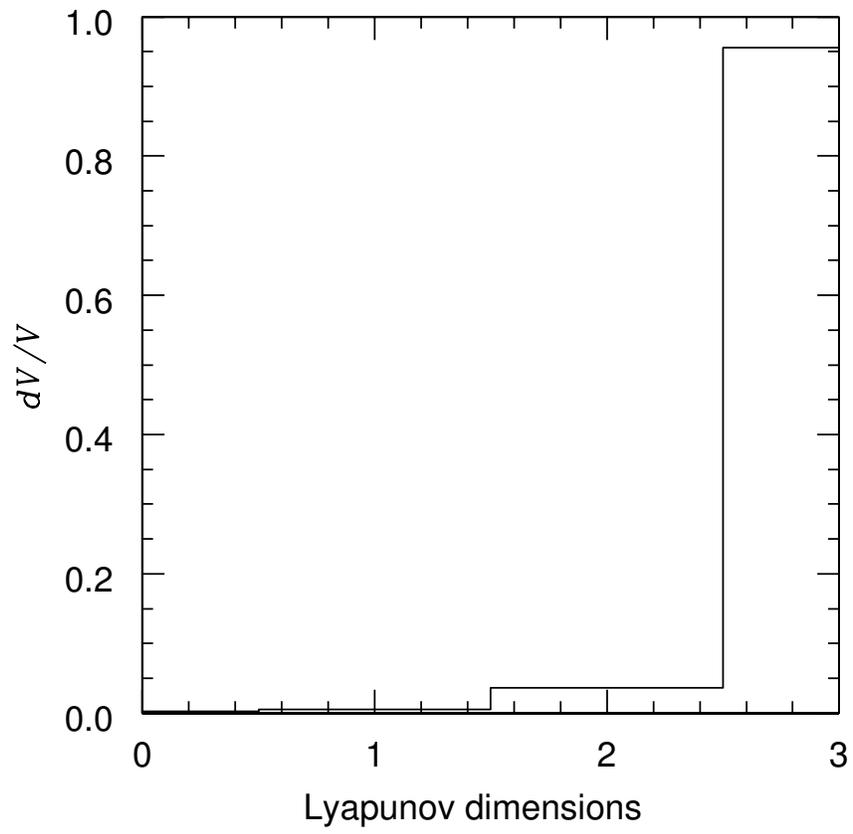

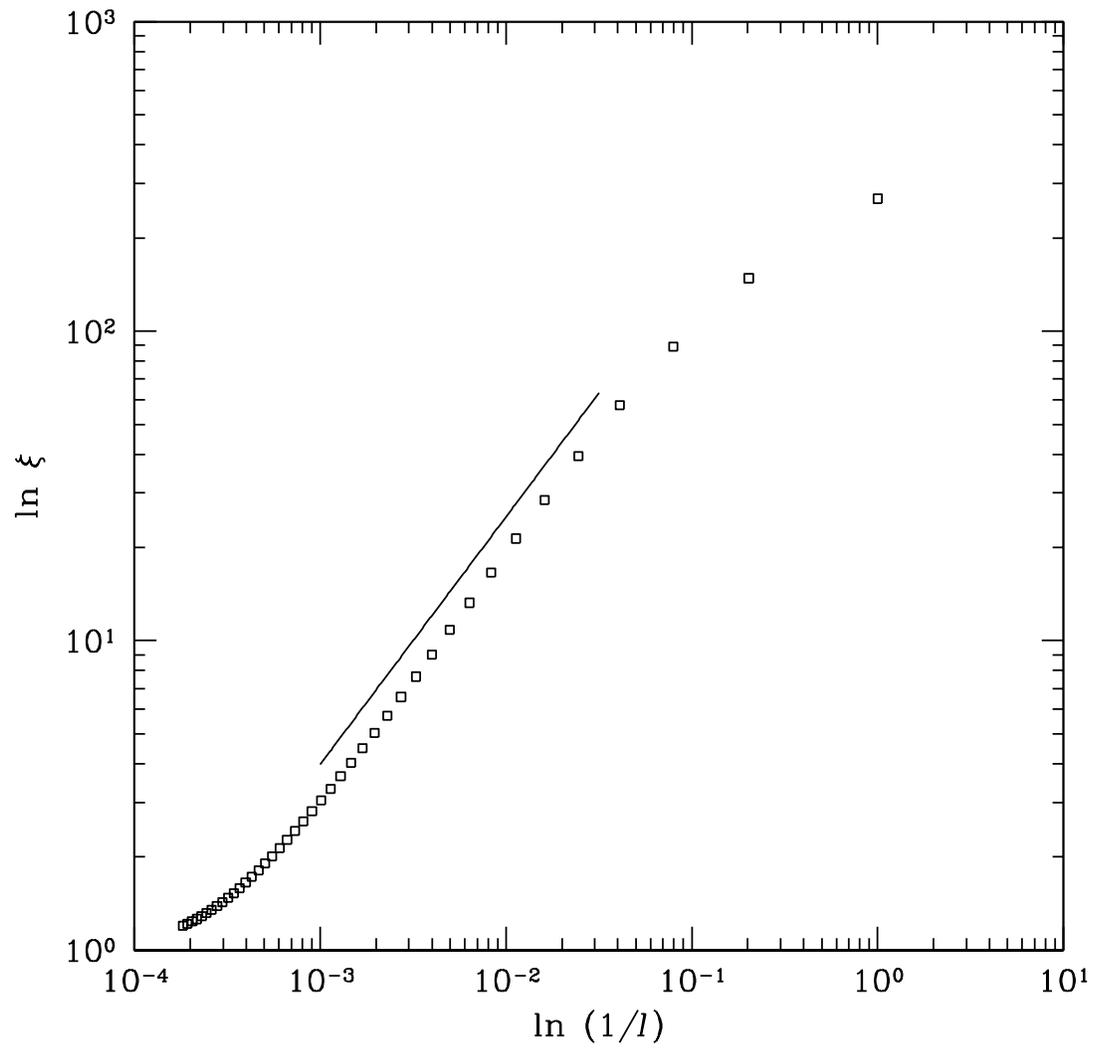